\documentclass[article]{jss-clean}

\usepackage{orcidlink, thumbpdf,lmodern}
\usepackage{framed}
\usepackage{amsmath, amsfonts, amssymb, booktabs}
\usepackage{comment}
\usepackage{appendix}
\usepackage{color}

\newcommand{\mv}[1]{{\boldsymbol{\mathrm{#1}}}}

\author{David Bolin~\orcidlink{0000-0003-2361-5465}\\
King Abdullah University of\\
Science and Technology
   \And Alexandre B. Simas~\orcidlink{0000-0003-2562-2829}\\King Abdullah University of \\
   Science and Technology}
\Plainauthor{David Bolin, Alexandre B. Simas}

\title{\pkg{rSPDE}: tools for statistical modeling using fractional SPDEs}
\Plaintitle{rSPDE: tools for statistical modeling using fractional SPDEs}
\Shorttitle{\pkg{rSPDE}: statistical modeling using fractional SPDEs}

\Abstract{
  The \proglang{R} software package \pkg{rSPDE} contains methods for approximating 
  Gaussian random fields based on fractional-order stochastic partial differential equations (SPDEs). 
  A common example of such fields are Whittle--Mat\'ern fields on bounded domains in $\mathbb{R}^d$, manifolds, or metric graphs. The package also implements various other models which are briefly introduced in this article. 
  Besides the approximation methods, the package contains methods for simulation, 
  prediction, and statistical inference for such models, as well as interfaces to 
  \pkg{INLA}, \pkg{inlabru} and \pkg{MetricGraph}. With these interfaces, fractional-order SPDEs can
  be used as model components in general latent Gaussian models, for which full Bayesian 
  inference can be performed, also for fractional models on metric graphs. 
  This includes estimation of the smoothness parameter 
  of the fields. This article describes the computational methods used 
  in the package and summarizes the theoretical basis for these.
  The main functions of the package are introduced, and their usage is illustrated through various examples. 
}

\Plainkeywords{Gaussian random fields, fractional-order partial differential equations,
Bayesian inference, spatial prediction, R}

\Address{
  David Bolin, Alexandre B Simas\\
  King Abdullah University of \\
  Science and Technology\\
  23955 Thuwal\\
  Saudi Arabia\\
  E-mail: \email{david.bolin@kaust.edu.sa}, \email{alexandre.simas@kaust.edu.sa}\\
  URL: \url{https://stochproc.kaust.edu.sa/}
}

\begin{document}

\section{Introduction} \label{sec:intro}
The stochastic partial differential equation (SPDE) approach introduced by \citet{lindgren11} is a popular method for computationally efficient inference based on Gaussian processes. 
It is based on the fact that a Gaussian process with a Mat\'ern covariance \citep{matern60} with parameters $\sigma, \kappa, \nu > 0$,
\begin{equation}\label{eq:materncov}
r(h) = \frac{\sigma^2}{2^{\nu-1}\Gamma(\nu)}(\kappa h) K_{\nu}(\kappa h),
\end{equation}
can be viewed as a solution to the SPDE
\begin{equation}\label{eq:spde}
(\kappa^2 - \Delta)^{\alpha/2} (\tau u) = \mathcal{W}, \quad \text{on $\mathcal{D}$},
\end{equation}
when $\mathcal{D} = \mathbb{R}^d$. Here $\Gamma(\cdot)$ is the gamma function, $K_{\nu}$ is a modified Bessel function of the second kind and of order $\nu$, $\Delta$ is the Laplacian, and $\mathcal{W}$ is Gaussian white noise. The parameters of the covariance function \eqref{eq:materncov} and the SPDE \eqref{eq:spde} are related through the expressions $\alpha = \nu + d/2$ and $\tau^2 = \sigma^{-2}\Gamma(\nu)\Gamma(\alpha)^{-1}(4\pi)^{-d/2}\kappa^{-2\nu}$.
\citet{lindgren11} proposed approximating the solution to \eqref{eq:spde} by restricting it to a bounded domain $\mathcal{D}$ and then using a finite element method (FEM) approximation to obtain a computationally efficient Gaussian Markov random field (GMRF) approximation of the solution. 
Another advantage of the method is that it facilitates various generalizations of the Gaussian Mat\'ern fields. One can introduce non-stationarity by allowing the parameters $\kappa$ and $\tau$ to be spatially varying functions, and in this case, the resulting models are typically referred to as generalized Whittle--Mat\'ern fields \citep{BK2020rational}. One can also formulate Mat\'ern-like random fields on other spatial domains $\mathcal{D}$, such as manifolds \citep{lindgren11} or metric graphs \citep{BSW2022}, by formulating the SPDE directly on that domain. 

These advantages have made the approach widely popular, see \citet{lindgren2022spde} for a recent overview. A reason for the success of the method is that it is implemented in the \pkg{INLA} \citep{inla} and \pkg{inlabru} \citep{inlabru} \proglang{R} \citep{Rref} packages, which makes the implementation easy for users even if they are not familiar with the theoretical details. However, a common criticism of the approach is that it requires $\alpha$ to be fixed to an integer value, which means that the smoothness of the random field is kept fixed during inference. This can be restrictive because a main reason for the popularity of the Mat\'ern covariance is that it allows for estimation of the smoothness from data, which is a crucial parameter for the quality of spatial predictions \citep{stein99}. 

In recent years, several attempts have been made to relax the assumption of integer values for $\alpha$. There are essentially two main approaches for this. The first combines a FEM approximation with a rational approximation of the differential operator in the SPDE \citep{BKK2020, BKK2018, BK2020rational}, whereas the second combines a FEM approximation with a rational approximation of the fractional power of the corresponding covariance operator $(\kappa^2 - \Delta)^{-\alpha}$ \citep{xiong2022}. The advantage of the first approach is that one then can obtain accurate approximations for a given sample of the SPDE. However, for statistical inference this is rarely important as the distributional properties are of most interest. Because of this, the covariance-based approach of \citet{xiong2022} is interesting as it provides good approximations of the covariance function while providing an approximation that is compatible with \pkg{INLA} and \pkg{inlabru}. Recently \citet{bolin2024linear} also showed that for stationary Gaussian processes with Mat\'ern covariance function on intervals, accurate and computationally efficient Markov approximations can be obtained without FEM approximation solely based on a rational approximation of the covariance operator.  

The \pkg{rSPDE} package implements the operator-based approximations of \citet{BK2020rational}, the covariance-based approximations of \citet{xiong2022}, and the rational approximations without FEM of \citet{bolin2024linear}. Besides the approximation methods, the package contains functions for using the approximations for sampling, prediction, and statistical inference. It also contains interfaces to \pkg{INLA} and \pkg{inlabru}, which means that the fractional-order SPDEs can be included in general latent Gaussian models that can be fitted to data using Bayesian methods. 
The package further contains an interface to the \pkg{MetricGraph} package \citep{metricgraph}, which enables the formulation of fractional-order SPDEs on metric graphs, such as street or river networks, which can be fitted to data using likelihood-based methods in \pkg{rSPDE} or using Bayesian methods through \pkg{INLA} or \pkg{inlabru}. Finally, the package implements various other SPDE-based models such as intrinsic Whittle--Mat\'ern fields, anisotropic Whittle--Mat\'ern fields \citep{fuglstad2015,hildeman2021deformed},  and non-separable spatio-temporal random fields with advection and diffusion terms \citep{lindgren2024spacetime,Clarotto2024}.

\pkg{rSPDE} is available from the Comprehensive \proglang{R} Archive Network (CRAN) at \url{https://cran.r-project.org/package=rSPDE}. A development version of the package is available at \url{https://github.com/davidbolin/rSPDE} and a homepage for the package with several vignettes explaining the various features of the package in more detail is available at \url{https://davidbolin.github.io/rSPDE/}.

The structure of the article is as follows. Section~\ref{sec:models} introduces the computational methods that are available in \pkg{rSPDE}. Section~\ref{sec:comparison} presents a comparison to illustrate the accuracy of the methods for Gaussian Mat\'ern fields and Section~\ref{sec:base} introduces the main functions of the package. Section~\ref{sec:inla} introduces the \pkg{INLA} interface, Section~\ref{sec:inlabru} the \pkg{inlabru} interface, and Section~\ref{sec:metricgraph} the \pkg{MetricGraph} interface. In each section, an illustrative example is used to show the capabilities of the package. Section~\ref{sec:othermodels} briefly introduces the various other models that are implemented in the package. Finally, future plans for the package are discussed in Section~\ref{sec:discussion}. 

\section{Computational methods} \label{sec:models}
In this section, we summarize the main theoretical ideas and approximation methods used in the \pkg{rSPDE} package. We start by discussing the generally applicable FEM-based approximation methods and then discuss the methods without FEM.

\subsection{FEM-based approximation methods}
Several popular Gaussian random field models can be represented as solutions to stochastic partial differential equations (SPDEs) of the form 
\begin{equation}\label{eq:Lspde}
L^{\beta} (\tau u) = \mathcal{W} \quad \text{on $\mathcal{D}$},
\end{equation}
Here $\mathcal{W}$ is Gaussian white noise, $L$ is a second-order differential operator, the fractional power $\beta>0$ determines the smoothness of $u$, and $\tau>0$ scales the variance of $u$. Examples include the generalized Whittle--Mat\'ern fields, obtained with $L = \kappa^2I - \Delta$, where $\kappa$ is a bounded function, which also is bounded away from $0$, the Whittle--Mat\'ern fields, which is the special case when $\kappa>0$ is constant, and the anisotropic Whittle-Mat\'ern fields with $L = I - \nabla\cdot(\boldsymbol{H}\nabla)$ where $\boldsymbol{H}$ is a symmetric and positive definite matrix. 

If $2\beta$ is an integer and if the domain $\mathcal{D}$ where the model is defined is bounded, then $u$ can be approximated by a Gaussian Markov random field (GMRF) $\boldsymbol{\mathrm{u}}$ via a finite element method (FEM) for the SPDE. Specifically, the approximation can be written as 
\begin{equation}\label{eq:uh}
u_h(s) = \sum_{i=1}^{n_h} u_i \varphi_i(s),
\end{equation}
were $\{\varphi_i\}$ are piecewise linear basis functions defined by some triangulation of $\mathcal{D}$ and the vector of weights $\boldsymbol{\mathrm{u}} = (u_1,\ldots,u_{n_h})^T$ is normally distributed, $N(\boldsymbol{\mathrm{u}},\tilde{\boldsymbol{\mathrm{Q}}}^{-1})$, where $\tilde{\boldsymbol{\mathrm{Q}}}$ is sparse, see \citet{lindgren11}.  

For a general $\beta>d/4$, the FEM approximation \eqref{eq:uh} solves the discrete SPDE
\begin{equation}\label{eq:Lh}
L_h^{\beta} (\tau u_h) = \mathcal{W}_h, 
\end{equation}
where $L_h$ is the FEM approximation of $L$ and $\mathcal{W}_h$ is Gaussian white noise on the space of piecewise linear functions on the mesh. There are two methods in \pkg{rSPDE} for obtaining a computationally efficient approximation of $u_h$ for general $\beta>d/4$. 

The first method is the operator-based rational approximation by \citet{BK2020rational}. This  combines the FEM approximation of \eqref{eq:Lspde} with a rational approximation of the fractional power $L^{-\beta}$ used to compute the solution $ \tau u_h = L_h^{-\beta}\mathcal{W}_h$. Specifically, one approximates $L_h^{-\beta}$ by 
$
L_{h,m}^{-\beta} = L_h^{-\max(\lfloor\beta\rfloor,1)} p(L_h^{-1})q(L_h^{-1})^{-1},
$
where $\lfloor\beta\rfloor$ denotes the integer part of $\beta$, $m$ is the order of rational approximation. Further, $p(L_h^{-1}) = \sum_{i=0}^m a_i L_h^{m-i}$ and $q(L_h^{-1}) = \sum_{j=0}^{m+1} b_j L_h^{m-i}$ are polynomials with coefficients $\{a_i\}_{i = 0}^m$ and $\{b_j\}_{j = 0}^{m+1}$ 
obtained from a rational approximation of the function $x^{\beta - \lfloor\beta\rfloor}$ on an interval that covers the spectrum of $L_h^{-1}$. 

This results in an approximation of the original SPDE which is of the form 
$P_l u_h = P_r \mathcal{W}_h$,
where $P_l$ and $P_r$ are non-fractional operators defined in terms of polynomials $p_l$ and $p_r$. The order of $p_r$ is given by $m$ and the order of $p_l$ is $m + m_{\beta}$ where $m_{\beta}$ is the integer part of $\beta$ if $\beta>1$ and $m_{\beta} = 1$ otherwise. The solution to this equation is an approximation $u_{h,m}$ of $u$ on the basis expansion form in \eqref{eq:uh}. 
The difference to the non-fractional case is that the vector of stochastic weights now is $\boldsymbol{\mathrm{u}} \sim N(\boldsymbol{\mathrm{0}},\boldsymbol{\mathrm{P}}_r\boldsymbol{\mathrm{Q}}^{-1}\boldsymbol{\mathrm{P}}_r^T)$ where $\boldsymbol{\mathrm{Q}}$ and $\boldsymbol{\mathrm{P}}_r$ are sparse matrices. Alternatively, $\boldsymbol{\mathrm{u}}$ can be represented as $\boldsymbol{\mathrm{u}} = \boldsymbol{P}_r\boldsymbol{\mathrm{x}}$ with 
$\boldsymbol{\mathrm{x}} \sim N(\boldsymbol{\mathrm{0}}, \boldsymbol{\mathrm{Q}}^{-1})$,
which means that the discrete approximation is a latent GMRF. 

The second type of rational approximation is the covariance-based approach introduced in \cite{xiong2022}. This is an efficient and more numerically stable alternative to the operator-based rational SPDE approach. The idea behind this approach is to use the fact that a centered Gaussian random field is uniquely specificed by its covariance operator. If we let $\beta = \alpha/2$, the solution $u_h$ to \eqref{eq:Lh} has a covariance operator is given by $L_h^{-\alpha}$. The covariance-based approximation directly approximates this fractional-order covariance operator instead of the corresponding differential operator.

The rational approximation is computed as 
$
L_{h,m}^{-\alpha} = L_h^{-\lfloor\alpha\rfloor} p(L_h^{-1})q(L_h^{-1})^{-1},
$
where $m$ again is the order of rational approximation, and 
$p(L_h^{-1}) = \sum_{i=0}^m a_i L_h^{m-i}$ and 
$q(L_h^{-1}) = \sum_{j=0}^m b_j L_h^{m-i}$ are polynomials with coefficients 
obtained from a rational approximation of $x^{\alpha - \lfloor\alpha\rfloor}$. 
Performing a partial fraction decomposition of the $p(L_h^{-1})q(L_h^{-1})^{-1}$ yields the representation 
$$
\mathbf{\Sigma}_{\textbf{u}} = (\textbf{L}^{-1}\textbf{C})^{\lfloor\alpha\rfloor} \sum_{i=1}^{m}r_i(\textbf{L}-p_i\textbf{C})^{-1}+k\textbf{K}, 
$$
for the covariance matrix of the stochastic weights $\textbf{u}$. Here, $k$ and $\{p_i, r_i\}_{i=1}^m$ are the coefficents of the partial fraction decomposition, satisfying $k, r_i>0$ and $p_i<0$ for $i=1,\ldots,m$. Further,$\textbf{C} = \{C_{ij}\}_{i,j=1}^{n_h}$ with $C_{ij} = (\varphi_i,\varphi_j)_{L_2(\mathcal{D})}$ is the mass matrix, $\textbf{L}$ is the matrix representation of $L_h$, and
$$\textbf{K}=\left\{
	\begin{array}{lcl}
		\textbf{C}      &      & {\lfloor\alpha\rfloor=0}\\
		\textbf{L}^{-1}(\textbf{C}\textbf{L}^{-1})^{\lfloor\alpha\rfloor-1}    &      & {\lfloor\alpha\rfloor\geq 1}\\
	\end{array} \right. .$$
	
This shows that we can write 
$
\hat{\textbf{u}}=\sum_{i=1}^{m+1}\textbf{x}_i,
$
where $\textbf{x}_i = (x_{i,1}, \ldots, x_{i,{n_h}}) \sim N(\textbf{0},\textbf{Q}_i^{-1}),$
with  
$$
\textbf{Q}_i=\left \{
	\begin{array}{lcl}
		(\textbf{L}-p_k\textbf{C})(\textbf{C}^{-1}\textbf{L})^{\lfloor\alpha\rfloor}/r_k,      &      & {i = 1,...,m}\\
		 (k\textbf{K})^{-1},   &      & {i = m+1}\\
	\end{array}. \right.
$$
Thus, the approximation can be represented as a sum of GMRFs. 

Both approximation methods thus lead to  GMRF representations which facilitate computationally efficient methods for inference and prediction. In both cases, one has to compute a rational approximation of the function $x^{\alpha}$ on an interval. The \pkg{rSPDE} package implements three different options for this task, described in Appendix~\ref{app:type_rational}. The type of approximation that is used has an effect on the quality of the approximation, but the choice is seldom of much importance for practical applications \citep{xiong2022}. 

\subsection{Markov approximations without FEM}
For Gaussian Mat\'ern fields on intervals, we can obtain computationally efficient Markov approximations without FEM, as introduced in \citet{bolin2024linear}. To see how, let $u$ be a centered Gaussian Process on an interval $I\subset \mathbb{R}$ with covariance function \eqref{eq:materncov}. If $\alpha\in\mathbb{N}$, this process has Markov properties which can be used for computationally efficient inference. 
Specifically, we have that $u$ is a Markov process of order $\alpha$, which is $\alpha-1$ times differentiable in the mean-squared sense. Because of this, the multivariate process $\mv{u}(t) = (u(t), u'(t),\ldots, u^{(\alpha-1)}(t))$ is a first order Markov process, and we can therefore use standard tools for Markov processes for computationally efficient inference and prediction. 

Suppose now that $\alpha\notin\mathbb{N}$. The process has spectral density 
$
f^{\alpha}(w) = A\sigma^2(\kappa^2 + w^2)^{-\alpha},
$
where $A=\frac{1}{2\pi}\Gamma(\alpha)\sqrt{4\pi}\kappa^{2\nu}\Gamma(\nu)^{-1}$. 
Performing the same type of rational approximation as we did for $L^{-\alpha}$ above, but on the spectral density, and then performing the partial fraction decomposition of $p(x)/q(x)$, we obtain
\begin{equation}\label{partial_frac}
 \begin{aligned}
f_m^{\alpha}(w)&= A\sigma^2\kappa^{-2\alpha}\left[\frac{k}{(1+\kappa^{-2}w^2)^{\lfloor \alpha \rfloor}}+\sum_{i=1}^{m}r_i\frac{1}{(1+\kappa^{-2}w^2)^{\lfloor \alpha \rfloor}(1+\kappa^{-2}w^2-p_i)}\right]\\
&=: f_{m,0}^{\alpha}(w)+\sum_{i=1}^{m}f_{m,i}^{\alpha}(w),
\end{aligned}
 \end{equation}
where $k,r_i>0$ and $p_i<0$ are the same coefficients as in the covariance-based rational approximation above. Based on this expression, one can compute the corresponding covariance function explicitly and show that this converges exponentially fast in the order $m$ to the true Mat\'ern covariance \citep{bolin2024linear}. 
Furthermore, because $f_m^{\alpha}(w)$ is a sum of valid spectral densities, a Gaussian process with this spectral density can be written as a sum of independent Gaussian processes
	$u(x)=u_0(x)+u_1(x)+\dotsc+u_m(x)$,
with $u_0$ has spectral density $f_{m,0}^{\alpha}$ and each $u_i$ has spectral density $f_{m,i}^{\alpha}$, $i=1,2,\dotsc,m$. These spectral densities are all reciprocals of polynomials, which means that each process $u_i, i=0,\ldots,m$ is a Gaussian Markov process \citep{Pitt1971, Rozanov1982Markov}. 
Specifically, we have that $u_0$ is a Markov process of order $\max(\lfloor\alpha\rfloor,1)$, which is $\max(\lfloor\alpha\rfloor-1,0)$ times differentiable in the mean-squared sense, and $u_i, i>0$, is a Markov process of order $\lceil\alpha\rceil$, which is $\lfloor\alpha\rfloor$ times differentiable in the mean-squared sense. Because of this, we can define the corresponding multivariate processes which are Markov of order one, just as in the case $\alpha\in\mathbb{N}$ and use this representation for computationally efficient inference, prediction and simulation.

\section{An illustration of the accuracy}\label{sec:comparison}
\begin{figure}[t!]
\centering
\includegraphics{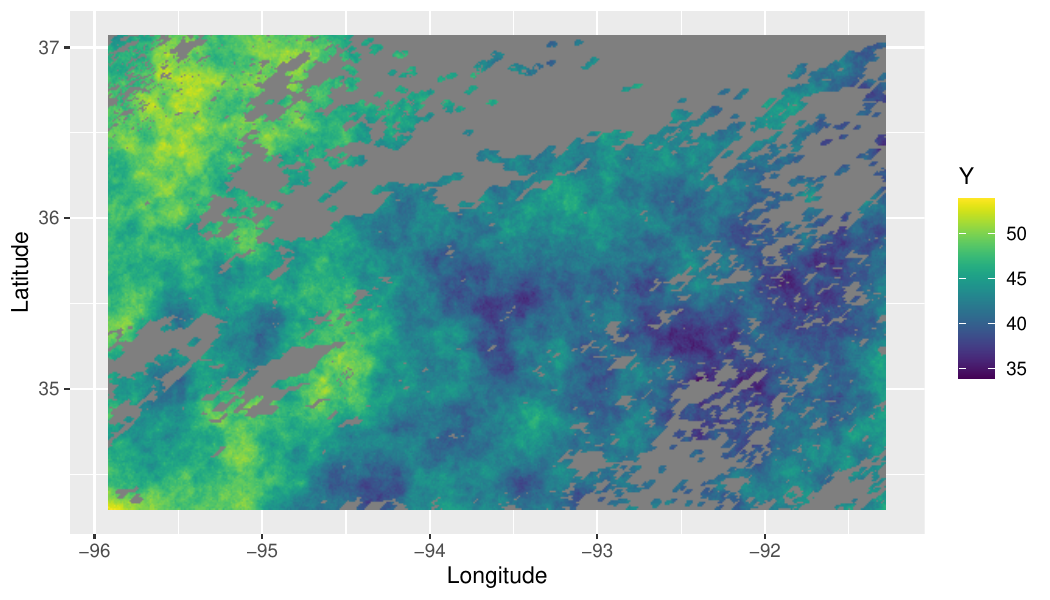}
\caption{Simulated data from \citet{heaton2019}.}
\label{fig:sim_data}
\end{figure}

As an illustration of how accurate the \pkg{rSPDE} methods are, we consider the simulated dataset used for the competition in \citet{heaton2019}. The dataset, shown in Figure~\ref{fig:sim_data}, was created by simulating a centered Gaussian process $u(s)$ with an exponential covariance function and creating the observations as $y_i = \beta + u(s_i) + \varepsilon_i$, where $\{\varepsilon_i\}$ are independent $N(0,\sigma_{\varepsilon}^2)$ variables. 

The goal of the challenge was to first estimate the parameters of this model (i.e., the nugget variance $\sigma_{\varepsilon}^2$, the intercept $\beta$, and the two parameters of the exponential covariance function) based on the data and then predict the values at the unobserved locations. 
The methods were compared in terms of the mean absolute error (MAE), the root-mean-squared error (RMSE), the continuous ranked probability score (CRPS), the interval score (INT) and the prediction interval coverage (CVG). All scores except CVG are negatively oriented so that a lower value is better, whereas the CVG has the target value 0.95. See \cite{heaton2019} for details. 

As an illustration, we use the covariance-based rational approximation method with different approximation orders $m=1,2,3$, using the same mesh as was used for the original SPDE approach in \cite{heaton2019}. The parameter estimation and the prediction were performed using the \pkg{inlabru} interface introduced in Section~\ref{sec:inlabru}, using the default values for all priors. 
The results are shown in Table~\ref{tab:comparison}, where the results for the first twelve methods are taken from \cite{heaton2019} and the final three \pkg{rSPDE} results are obtained using the code in Appendix~\ref{app:comparison}. We can note that the results for the \pkg{rSPDE} method gets more accurate as $m$ increases, and when $m=3$, the method has the best scores in all metrics.

\begin{table}[t]
\centering
\caption{Numerical scores the simulated data from \cite{heaton2019}. The results for the first 12 methods are taken from the original article.}
\label{tab:comparison}
 \small 
\begin{tabular} {lllllll}
\toprule
Method            & MAE  & RMSE & CRPS & INT & CVG \\
\hline
FRK               & 1.03 & 1.31 & 0.74 & 8.35 & 0.84\\
Gapfill             & 0.73 & 1.00 & 0.64 & 18.01 & 0.44\\
LatticeKrig         & 0.63 & 0.87 & 0.45 & 4.04 & 0.97\\
LAGP                & 0.79 & 1.11 & 0.57 & 5.71 & 0.90\\
Metakriging         & 0.74 & 0.97 & 0.53 & 4.69 & 0.99\\
MRA                 & 0.61 & 0.83 & 0.43 & 3.64 & 0.93\\
NNGP                & 0.65 & 0.88 & 0.46 & 3.79 & 0.96\\
Partition           & 0.64 & 0.86 & 0.47 & 5.05 & 0.86\\
Pred.~Proc.         & 1.06 & 1.43 & 0.76 & 7.33 & 0.89\\
Tapering            & 0.69 & 0.97 & 0.55 & 6.39 & 1.00\\
Periodic Embedding  & 0.65 & 0.91 & 0.47 & 4.16 & 0.97\\
SPDE                & 0.62 & 0.86 & 0.59 & 7.81 & 1.00\\
\hline
rSPDE $m=1$         & 0.61 & 0.82 & 0.43 & 4.01 & 0.91\\
rSPDE $m=2$         & 0.60 & 0.82 & 0.43 & 3.61 & 0.94\\
rSPDE $m=3$         & 0.60 & 0.81 & 0.42 & 3.53 & 0.95\\
\bottomrule
\end{tabular}
\label{tab:sim_res}
\end{table}

\section[Base rSPDE methods]{Base \pkg{rSPDE} methods}\label{sec:base}
In this section, we introduce the main methods of the \pkg{rSPDE} package which are not meant to be used in combination with other packages. In later sections, we introduce the interfaces to \pkg{INLA}, \pkg{inlabru} and \pkg{MetricGraph}.

\subsection{Constructing FEM-based approximations}
There are three main functions for defining FEM-based rational approximations of \eqref{eq:Lh}. The most general function is \code{fractional.operators}, which works for a wide class of models with a general differential operator $L$. For the stationary Matérn case, where $L = \kappa^2 - \Delta$, the function \code{matern.operators} provides a simplified model specification. For the generalized non-stationary Matérn model, defined through the SPDE \eqref{eq:spde} where $\kappa$ and $\tau$ are functions, the function \code{spde.matern.operators} can be used. For univariate domains, approximations without FEM can be constructed using the \code{matern.rational} function. 

Let us illustrate how these functions can be used to construct an approximation of a Whittle--Mat\'ern field on the unit interval $[0, 1]$. 
The first step for constructing the FEM-based rational SPDE approximation is to define the FEM mesh. In this section, we use the simple FEM implementation in the \code{rSPDE} package for models defined on an interval. 
We then start by defining a vector with mesh nodes $s_i$ where the basis functions $\varphi_i$ are centered. 
\begin{Schunk}
\begin{Sinput}
R> s <- seq(from = 0, to = 1, length.out = 101)
\end{Sinput}
\end{Schunk}

We can now use \code{matern.operators} to construct a rational approximation of order $m=1$ for a Gaussian random field with a Matérn covariance function on the interval. The model can be specified either using the SPDE parameters $\kappa,\alpha,\tau$ or the covariance parameters $r,\nu,\sigma$, where $r = \sqrt{8\nu}/\kappa$ is the practical correlation range. 
For an operator-based approximation, we must set \code{type="operator"} in the function.
\begin{Schunk}
\begin{Sinput}
R> par <- list(sigma = 2, nu = 0.8, r = 0.15, kappa = sqrt(8*0.8)/0.15)
R> op1 <- matern.operators(sigma = par$sigma, range = par$r, nu = par$nu, 
+                          loc_mesh = s, d = 1, m = 1, type = "operator", 
+                          parameterization = "matern")
R> op2 <- matern.operators(sigma = par$sigma, range = par$r, 
+                          nu = par$nu, loc_mesh = s, d = 1, m = 1, 
+                          parameterization = "matern")
\end{Sinput}
\end{Schunk}
The object \code{op1} contains the matrices needed for evaluating the distribution of the stochastic weights $\boldsymbol{\mathrm{u}}$ for the operator-based approximation, and \code{op2} contains the corresponding matrices for a covariance-based approximation. 

By specifying \code{loc_mesh}, \code{matern.operators} assembles the required finite element matrices internally, i.e., the mass matrix $\boldsymbol{\mathrm{C}}$ and the stiffness matrix $\boldsymbol{\mathrm{G}}$, with elements $G_{ij} = \int \nabla\varphi_j(s) \cdot \nabla\varphi_i(s) ds$.  These can also be constructed manually though the function \code{rSPDE.fem1d}. 

If we want to evaluate $u_h(s)$ at some locations $s_1,\ldots, s_n$, we need to multiply the weights with the basis functions $\varphi_i(s)$ evaluated at the locations. For this, we can construct the observation matrix $\boldsymbol{\mathrm{A}}$ with elements $A_{ij} = \varphi_j(s_i)$, which links the FEM basis functions to the locations. This matrix can be constructed using the function \code{rSPDE.A1d}.

\begin{figure}[t!]
\centering
\includegraphics{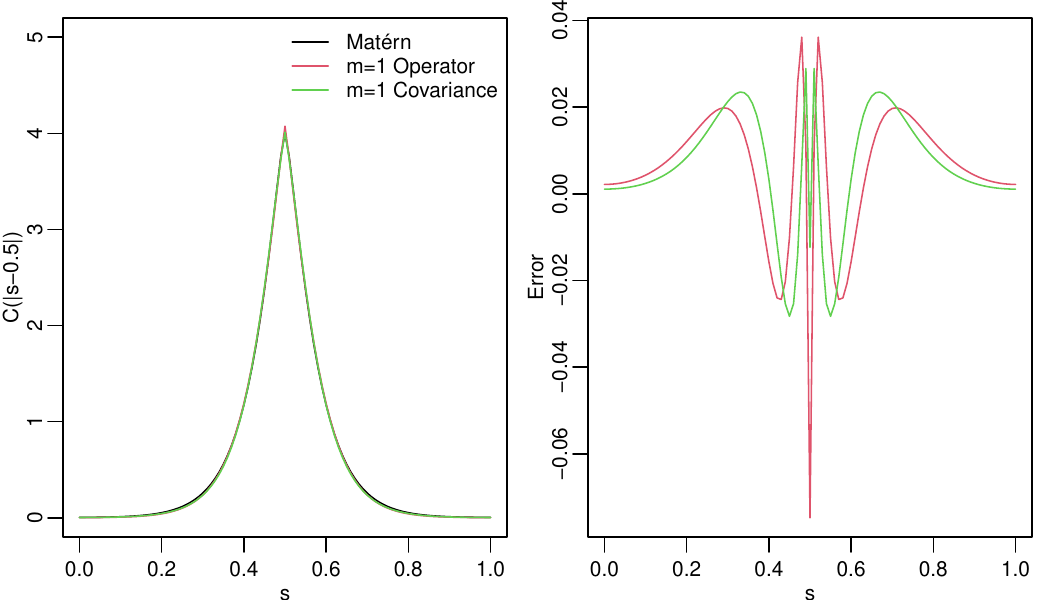}
\caption{\label{fig:op1} True Rational Mat\'ern covariance function and rational approximations (left) and errors for the approximations (right).}
\end{figure}

To evaluate the accuracy of the approximation, let us compute the covariance function between the process at $s=0.5$ and all other locations in \code{s} and compare with the true covariance function, which is the folded Matérn covariance, see Theorem 1 in \citet{lindgren11}. The covariances can be computed through the function \code{cov_function_mesh} in the model object. 
\begin{Schunk}
\begin{Sinput}
R> c_op <- op1$cov_function_mesh(0.5, direct = TRUE)
R> c_cov <- op2$cov_function_mesh(0.5)
R> c_true <- folded.matern.covariance.1d(rep(0.5, length(s)), abs(s), 
+                                        par$kappa, par$nu, par$sigma)
\end{Sinput}
\end{Schunk}
The covariance function and the error compared with the Mat\'ern covariance are shown in Figure~\ref{fig:op1}. The argument \code{direct = TRUE} specifies that the operator-based covariance is calculated as 
$
\boldsymbol{\mathrm{P}}_r \boldsymbol{\mathrm{Q}}^{-1}\boldsymbol{\mathrm{P}}_r^T\boldsymbol{\mathrm{v}},
$
where  $\boldsymbol{\mathrm{v}}$ is a vector with all basis functions evaluated in $s=0.5$. 
This may be problematic in some cases due to large condition numbers of the matrices involved. 
To handle such issues, the package contains functions for performing operations such as $\boldsymbol{\mathrm{P}}_rv$ or $\boldsymbol{\mathrm{P}}_r^{-1}v$ that takes advantage of the structure of $\boldsymbol{\mathrm{P}}_r$ to avoid numerical instabilities. A complete list of these function can be seen by typing \code{?operator.operations}, and these are used by default unless manually turned off (e.g., by specifying \code{direct=TRUE} in \code{cov_function_mesh}).

To improve the approximation we can increase the degree of the polynomials, by increasing $m$, and/or increasing the number of basis functions used for the FEM approximation. 
Since the error induced by the rational approximation decreases exponentially in $m$, there is rarely a need for an approximation with a large value of $m$. This is advantageous because the computational cost of both the operator-based and covariance-based approaches increases with $m$. 

Let us, as an example, compute the approximations with a sligthly finer mesh for $m=1,\ldots,4$. To compare the error on the original mesh, we load the \pkg{fmesher} package \citep{fmesher} to use the \code{fm_basis} and \code{fm_mesh_1d} functions to map between the meshes.
For the operator-based version, we also compute the error with and without using the methods for handling numerical instabilities.

\begin{Schunk}
\begin{Sinput}
R> s2 <- seq(from = 0, to = 1, length.out = 501)
R> A  <- fm_basis(fm_mesh_1d(s2), s)
R> err_op <- err_op2 <- err_cov <- rep(0, 4)
R> op <- list()
R> for (i in 1:4) {
+      op[[i]] <- matern.operators(range = par$r, sigma = par$sigma, 
+                                  nu = par$nu, loc_mesh = s2, d = 1, 
+                                  m = i, type = "operator",
+                                  parameterization = "matern")
+      c_op <- A %*% op[[i]]$cov_function_mesh(0.5, direct = TRUE)
+      err_op[i] <- norm(c_true - c_op)
+      err_op2[i] <- norm(c_true - A %*% op[[i]]$cov_function_mesh(0.5))
+      op_cov <- matern.operators(range = par$r, sigma = par$sigma, 
+                                 nu = par$nu, loc_mesh = s2, d = 1, 
+                                 m = i, parameterization = "matern")
+      err_cov[i] <- norm(c_true - A %*% op_cov$cov_function_mesh(0.5))
+  }
R> print(t(data.frame(operator = err_op, operator.stable = err_op2,
+                     covariance = err_cov, 
+                     row.names = c("m=1","m=2","m=3","m=4"))))
\end{Sinput}
\begin{Soutput}
                     m=1        m=2          m=3          m=4
operator        1.185747 0.84251825 2.860902e+03 64.805252303
operator.stable 1.185748 0.11738832 2.339077e-02  0.018882323
covariance      1.172898 0.09956936 1.805060e-02  0.007863185
\end{Soutput}
\end{Schunk}

We indeed see that the operator-based approximation which is not using the numerically stable matrix calculations is numerically unstable for $m>2$. These issues are not present when using the stable methods or when using the covariance-based approximation. Because of the greater numerical stability of the covariance-based approximation, we recommend using this whenever working with Gaussian processes. The covariance-based approximation is also the only option that is compatible with \pkg{INLA} and \pkg{inlabru}. One situation where the operator-based approach is needed is when working with non-Gaussian fields such as those in \citet{bolin2020typeg}. We will, however, not go into details about that here.

Let us now examine a non-stationary model with $\kappa(s) = 10(1+2s^2)$ and $\tau(s) = 0.1(1 - 0.7s^2)$. We can then use \code{spde.matern.operators} to create a rational approximation as follows.
\begin{Schunk}
\begin{Sinput}
R> s_mesh <- fm_mesh_1d(s2)
R> kappa_ns <- 10 * (1 + 2 * s2^2)
R> tau_ns <- 0.1 * (1 - 0.7 * s2^2)
R> op <- spde.matern.operators(kappa = kappa_ns, tau = tau_ns, 
+                              nu = 0.8, d = 1, m = 1, 
+                              mesh = s_mesh, type = "operator", 
+                              parameterization = "matern")
\end{Sinput}
\end{Schunk}
Let us compute the covariance function $C(s,s_i)$ of the non-stationary model for the locations $s_1=0.1, s_2 = 0.5,$ and $s_3 = 0.9$. 
\begin{Schunk}
\begin{Sinput}
R> v <- t(op$make_A(c(0.1, 0.5, 0.9)))
R> covs <- Sigma.mult(op, v)
\end{Sinput}
\end{Schunk}

\begin{figure}[t!]
\centering
\includegraphics{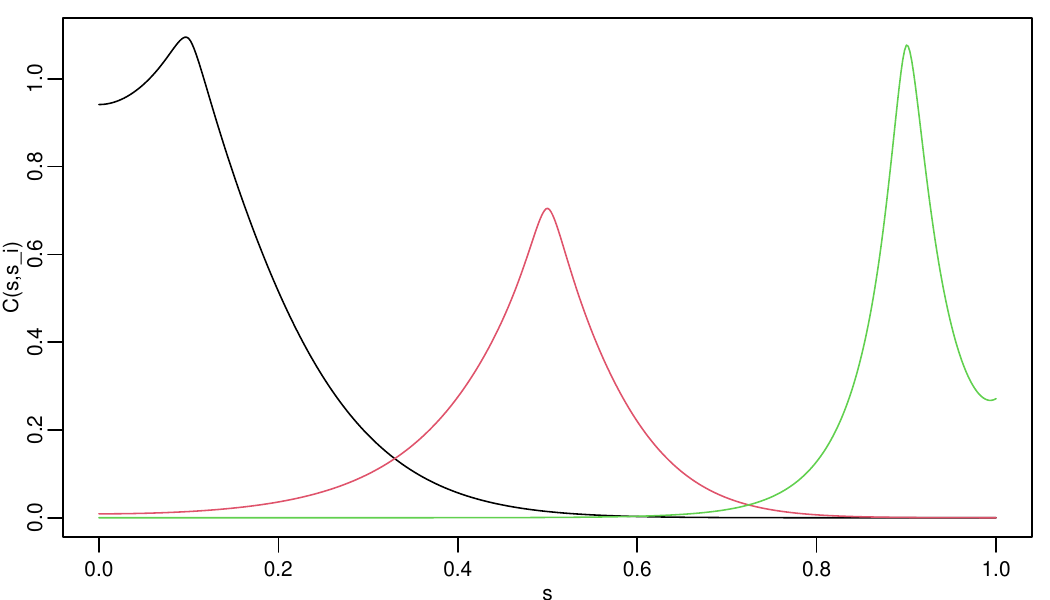}
\caption{\label{fig:op3} Non-stationary covariances.}
\end{figure}

The three covariances are shown in Figure~\ref{fig:op3}. We see that this choice of $\kappa(s)$ and $\tau(s)$ results in a model with longer range for small values of $s$ and smaller variance in the middle of the domain. 

We can also apply the general function \code{fractional.operators} to construct the approximation. This function requires that the user supplies a discretization of the non-fractional operator $L$, as well as a scaling factor $c>0$ which is a lower bound for the smallest eigenvalue of $L$. In our case we have $L = \kappa(s)^2 - \Delta$, and the eigenvalues of this operator is bounded from below by $c = \min_s \kappa(s)^2$. We compute this constant and the discrete operator. 
\begin{Schunk}
\begin{Sinput}
R> fem <- fm_fem(s_mesh)
R> L <- fem$g1  + fem$c0 %*% Diagonal(501, kappa_ns^2)
\end{Sinput}
\end{Schunk}
Another difference between \code{fractional.operators} and the previous functions for constructing the approximation, is that it requires specifying $\beta$ instead of the smoothness parameter $\nu$ for the Matérn covariance. These two parameters are related as $2\beta = \nu + d/2$. 
\begin{Schunk}
\begin{Sinput}
R> op <- fractional.operators(L = L, beta = (0.8 + 1 / 2) / 2, 
+                             C = fem$c0, scale.factor = min(kappa_ns)^2, 
+                             tau = tau_ns, m = 1)
\end{Sinput}
\end{Schunk}
Let us make sure that we have the same approximation by comparing with the previously computed covariances.
\begin{Schunk}
\begin{Sinput}
R> norm(covs - Sigma.mult(op, v))
\end{Sinput}
\begin{Soutput}
[1] 0
\end{Soutput}
\end{Schunk}
Obviously, it is simpler to use \code{spde.matern.operators} in this case, but the advantage with \code{fractional.operators} is that it also can be used for other more general models such as one with $L = \kappa(s)^2 - \nabla \cdot (\boldsymbol{\mathrm{H}}(s) \nabla)$ for some matrix-valued function $\boldsymbol{\mathrm{H}}(s)$. 

\subsection{Constructing approximations without FEM}
The construction of the approximations without FEM is done in essentially the same way as above. 
Assume that we want to define a model on the interval $[0,1]$, which we want to evaluate at the locations \code{s} defined above. We can now use \code{matern.rational} to construct a rational SPDE approximation of a Gaussian random field with a Matérn covariance function on the interval. 
The object returned by this function contains the information needed for evaluating the approximation. Note, however, that the approximation is invariant to the locations \code{loc}, and they are only supplied to indicate where we want to evaluate it. 

To evaluate the accuracy of these types of approximations, let us compute the covariance function between the process at $s=0$ and all other locations in \code{s} and compare with the true Matérn covariance function. The covariances can be calculated by using the \code{covariance} method of the operator object, and we do this for $m=1,\ldots,4$.

\begin{Schunk}
\begin{Sinput}
R> c_true <- matern.covariance(abs(s[1] - s), par$kappa, par$nu, par$sigma)
R> for (i in 1:4) {
+    op_i <- matern.rational(loc = s, range = par$r, sigma = par$sigma, 
+                                nu = par$nu, m = i)
+    err_op[i] <- norm(c_true - op_i$covariance(ind = 1))
+  }
R> print(err_op)
\end{Sinput}
\begin{Soutput}
[1] 1.62287199 0.34501949 0.09382026 0.03064511
\end{Soutput}
\end{Schunk}
As expected, we see that the error decreases rapidly when we increase $m$ from $1$ to $4$.


\subsection{Simulation, inference and prediction}
Any \pkg{rSPDE} model, for example constructed via \code{fractional.operators}, \code{matern.operators}, \code{spde.matern.operators}, or \code{matern.rational}, can be simulated using the \code{simulate} method and fitted to data using the \code{rspde_lme} function. To illustrate the \code{simulate} method, let us simulate data from a non-stationary fractional model on the sphere. For this, we use \pkg{fmesher} to define a mesh on the sphere.
\begin{Schunk}
\begin{Sinput}
R> mesh <- fm_rcdt_2d(globe = 40)
\end{Sinput}
\end{Schunk}
We now define a fractional SPDE model on the mesh using the \code{rspde.matern} function, where $\kappa$ and $\tau$ satisfy the log-linear regressions $\log(\kappa(\mathbf{s})) = \theta_1 + \theta_2 b(\mathbf{s})$ and $\log(\tau(\mathbf{s})) = \theta_1 + \theta_3 b(\mathbf{s})$
where $b(\mathbf{s})$ is the latitude. 
We follow the same structure as the basic SPDE models in \pkg{INLA} when specifying the non-stationary parameters. Specifically, the covariates in the log-linear regressions are specified through the matrices \code{B.tau} and \code{B.kappa}. The columns of these matrices correspond to the same parameter. The first column does not have any parameter to be estimated. So, for instance, if one wants to share a parameter between \code{kappa} and \code{tau} one simply lets the corresponding column to be nonzero on both \code{B.kappa} and \code{B.tau}.
In \pkg{rSPDE} one can alternatively also specify log-linear regression on the (approximate) standard deviations and practical correlation ranges through the matrices \code{B.sigma} and \code{B.range}. 

\begin{figure}[t!]
\centering
\includegraphics[width = 0.45\linewidth]{"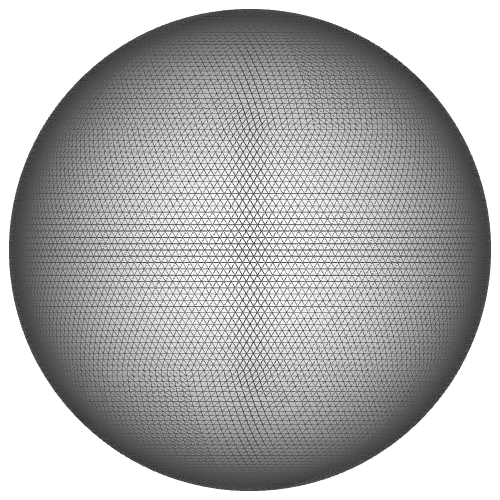"}
\includegraphics[width = 0.45\linewidth]{"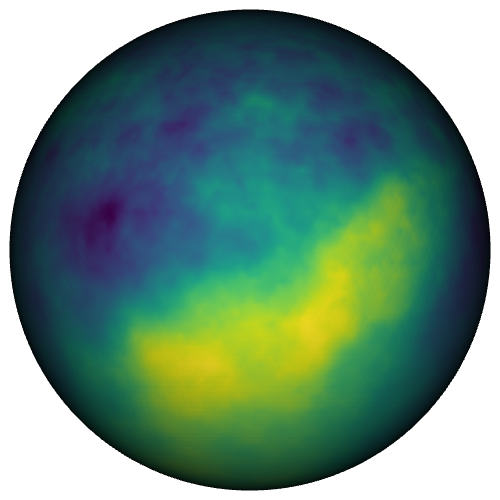"}
\caption{\label{fig:sim1} FEM mesh and simulation on the sphere.}
\end{figure}

We assume $\alpha = 2.3$ and $\theta_1 = 0, \theta_2 = 4$ and $\theta_3=-1$. Let us now build the model with the \code{spde.matern.operators} function:

\begin{Schunk}
\begin{Sinput}
R> theta <- c(0,4,-1)
R> Bkappa = cbind(0, 1,  mesh$loc[,3], 0)
R> Btau   = cbind(0, 1, 0, mesh$loc[,3])
R> op <- spde.matern.operators(mesh = mesh, theta = theta, alpha = 2.3, 
+                                     B.kappa = Bkappa, B.tau = Btau)
\end{Sinput}
\end{Schunk}
We can now simulate from the model as follows:
\begin{Schunk}
\begin{Sinput}
R> u <- simulate(op, seed = 10)
\end{Sinput}
\end{Schunk}
The results can be seen in Figure~\ref{fig:sim1}, which are plotted using \pkg{rgl} \citep{rgl} and \pkg{fmesher}. 

Let us generate a few replicates of the field and then generate some data from these fields, which we assume are observed at $m$ locations, $\{\mathbf{s}_1 , \ldots , \mathbf{s}_m \}$.
For each $i = 1,\ldots,m,$ we have
$$
y_{ij} = u_j(\mathbf{s}_i)+\varepsilon_{ij},
$$
where $u_j$ denotes the $j$th simulation and $\varepsilon_{ij}$ are independent cenetered Gaussian variables with standard deviation 0.1.

\begin{Schunk}
\begin{Sinput}
R> n_rep <- 50
R> m <- 1000
R> u_rep <- simulate(op, nsim = n_rep, seed = 10)
R> loc_mesh <- mesh$loc[sample(mesh$n,m),]
R> A <- fm_basis(x = mesh, loc = loc_mesh)
R> sigma_e <- 0.1
R> Y_rep <- A %*% u_rep + sigma_e * matrix(rnorm(m * n_rep), ncol = n_rep)
\end{Sinput}
\end{Schunk}
Note that \code{Y_rep} is a matrix with \code{n_rep} columns, each column containing one replicate. We now create an auxiliary vector repl indexing the replicates of y and store the data in a data frame which also includes the observation locations

\begin{Schunk}
\begin{Sinput}
R> repl <- rep(1:n_rep, each = m)
R> df_data_ns <- data.frame(y = as.vector(Y_rep), repl = repl,
+                           x_coord = rep(loc_mesh[,1], n_rep), 
+                           y_coord = rep(loc_mesh[,2], n_rep),
+                           z_coord = rep(loc_mesh[,3], n_rep))
\end{Sinput}
\end{Schunk}

We are now ready to fit the model to the data using the \code{rspde_lme} function, which provides maximum likelihood estimation of linear mixed effects models of the form considered here, which possibly have fixed effects for the mean value and a Gaussian process to capture the spatial dependence. This function has a \code{formula} argument which describes the relation between the response variable and possible fixed effects. In this case, we have no fixed effects, so we will simply use \code{formula = y ~ -1} to specify that we are fitting a centered Gaussian field. We then need to specify the model for the Gaussian process to fit, for this we create a new model object, using a coarser mesh to emulate the fact that we rarely have data that can be assumed to come from the discretized model. 
\begin{Schunk}
\begin{Sinput}
R> mesh <- fm_rcdt_2d(globe = 20)
R> op <- spde.matern.operators(mesh = mesh, B.kappa = Bkappa, B.tau = Btau)
\end{Sinput}
\end{Schunk}
Here we provide the matrices with the covariates for the non-stationary parameters, but 
we do not specify the parameters or the smoothness. In this way, the parameters including the smoothness will be estimated from data. If we would provide \code{nu} when defining the model, that parameter would be kept fixed throughout the estimation. 
The remaining arguments required for \code{rspde_lme} are the data frame, the names of the spatial coordinates in the data frame, and the replicate vector. We further set the argument \code{parallel} to \code{TRUE} which means that the numerical optimization of the likelihood will be done using the \pkg{optimParallel} \citep{optimParallel} package to speed up the inference. Various other options can be set, such as starting values of the parameter estimates, and to illustrate this, we specify the starting values of the latent field 
\begin{Schunk}
\begin{Sinput}
R> fit_ns <- rspde_lme(y ~ -1, model = op, data = df_data_ns, 
+                      repl = repl, parallel = TRUE,
+                      starting_values_latent = theta,
+                      loc = c("x_coord", "y_coord", "z_coord"))
\end{Sinput}
\end{Schunk}
The function also computes standard errors of the estimates based on numerical approximations of the hessian matrix of the estimates. The result object has a summary method which provides similar information as one would get if a model was fitted using the \code{lme} function of the \pkg{lme4} package \cite{lme4}. Having estimated a model using \code{rspde_lme}, one can use it for prediction using the \code{augment} function. For this, we first create a data frame with the locations where we want to predict
\begin{Schunk}
\begin{Sinput}
R> m_pred <- fm_rcdt_2d(globe = 11)
R> df_pred <- data.frame(x = m_pred$loc[,1], y = m_pred$loc[,2], 
+                        z = m_pred$loc[,3])
R> pred <- predict(fit_ns, newdata = df_pred, 
+                  loc = c("x","y","z"), which_repl = 3)
\end{Sinput}
\end{Schunk}
The predicted values are now stored in \code{pred$mean}. 
Alternatively, one can use the corresponding \code{augment} method. The difference between the two is that \code{augment} augments the supplied dataset with the predictions, residuals and standard errors for the fitted
values, whereas predict only computes the prediction. 

\section[INLA interface]{\pkg{INLA} interface}\label{sec:inla}

The \pkg{rSPDE} package has an interface to \pkg{INLA} which allows the models mentioned above to be included as components in general latent Gaussian models that can be fitted to data using \pkg{INLA}. 
We illustrate this through an application to a data set that consists of precipitation measurements 
from the Paraná region in Brazil. However, first we give a brief introduction to the main functions. 

\subsection{Overview of the main functions and options}
The \pkg{rSPDE} implementation is by design 
very similar to the implementation of SPDE models in \pkg{INLA}, so its usage should be
straightforward for \pkg{INLA} users. Table~\ref{tab:inla} shows the standard functions
which are used for SPDE models in \pkg{INLA} and the corresponding function in \pkg{rSPDE}. The main differences when comparing the arguments between the \pkg{rSPDE} implementation
and the standard SPDE implementation in \pkg{INLA} are the \code{nu} and \code{rspde.order} arguments,which are present in the \pkg{rSPDE} functions but not in the corresponding \pkg{INLA} functions. 

\begin{table}[t!]
\centering
\begin{tabular}{lll}
\hline
& \pkg{INLA} function           & \pkg{rSPDE} function\\ 
Model creation & \code{inla.spde2.matern} & \code{rspde.matern} \\
Index creation & \code{inla.spde.make.index} & \code{rspde.make.index} \\ 
Observation matrix & \code{inla.spde.make.A} & \code{rspde.make.A} \\
\hline

\end{tabular}
\caption{\label{tab:inla} Overview of functions used in \pkg{INLA} when creating SPDE models and the corresponding function in \pkg{rSPDE}.}
\end{table}

Specifying \code{rspde.order} determines the order of the rational approximation (i.e., the value of $m$ in the rational approximation). The default order is 1, and increasing this value will result in a more accurate and more computationally expensive approximation. Importantly, if a non-default value is used, this must be set in all function in Table~\ref{tab:inla}. 

Specifying \code{nu} indicates that the model has a fixed smoothness, given by the value specified. If we fix $\nu$ so that $\alpha = \nu + d/2$ is an integer in \code{rspde.matern}, then we must provide $\nu$ also in \code{rspde.make.index} and \code{rspde.make.A}. If we, on the other hand fix $\nu$ to a value so that $\alpha$ is not an integer, there is no need in providing $\nu$ in \code{rspde.make.index} and \code{rspde.make.A}. However, to avoid errors, we suggest providing $\nu$ in all function in Table~\ref{tab:inla} if $\nu$ should be kept fixed. 

If \code{nu} is not kept fixed, we need to provide an upper bound for it when creating the model in 
\code{rspde.matern}. The reason being that the sparsity of the precision matrix must be fixed during the estimation in \pkg{INLA},  and the higher the value of $\nu$ the denser the precision matrix is. 
This means that the higher the value of $\nu$, the higher the computational cost
to fit the model. Therefore, ideally, want to choose an upper bound for $\nu$
as small as possible, but larger than the ``correct'' value of $\nu$. 
The default value of the upper bound is 2, and to change this value, the argument \code{nu.upper.bound} can be used.

Which priors to use for the parameters are also specified in \code{rspde.matern}, and we provide details about this in Appendix~\ref{sec:inladetails}. The appendix also contains details on how to set the starting values for the parameters and how to change the type of rational approximation.

\subsection{Application}
The data consist of precipitation measurements from the Paraná region in Brazil
and were provided by the Brazilian National Water Agency. The data were collected
at 616 gauge stations in Paraná state, south of Brazil, for each day in 2011, and has also been used in \citet{krainski2018advanced}.  

We begin by loading the data and the boarder of the region from the \pkg{INLA} package:

\begin{Schunk}
\begin{Sinput}
R> data(PRprec)
R> data(PRborder)
\end{Sinput}
\end{Schunk}

The data frame contains daily measurements at 616 stations for the year 2011, as well as coordinates and altitude information for the measurement stations. We will not analyze the full spatio-temporal data set, but instead follow \citet{bolinlindstrom} look at the total precipitation in January, which we calculate as follows. We then extract the coordinates and remove the locations with missing values.

\begin{Schunk}
\begin{Sinput}
R> Y <- rowMeans(PRprec[, 3 + 1:31])
R> ind <- !is.na(Y)
R> Y <- Y[ind]
R> loc <- as.matrix(PRprec[ind, 1:2])
\end{Sinput}
\end{Schunk}

\begin{figure}[t!]
\centering
\includegraphics{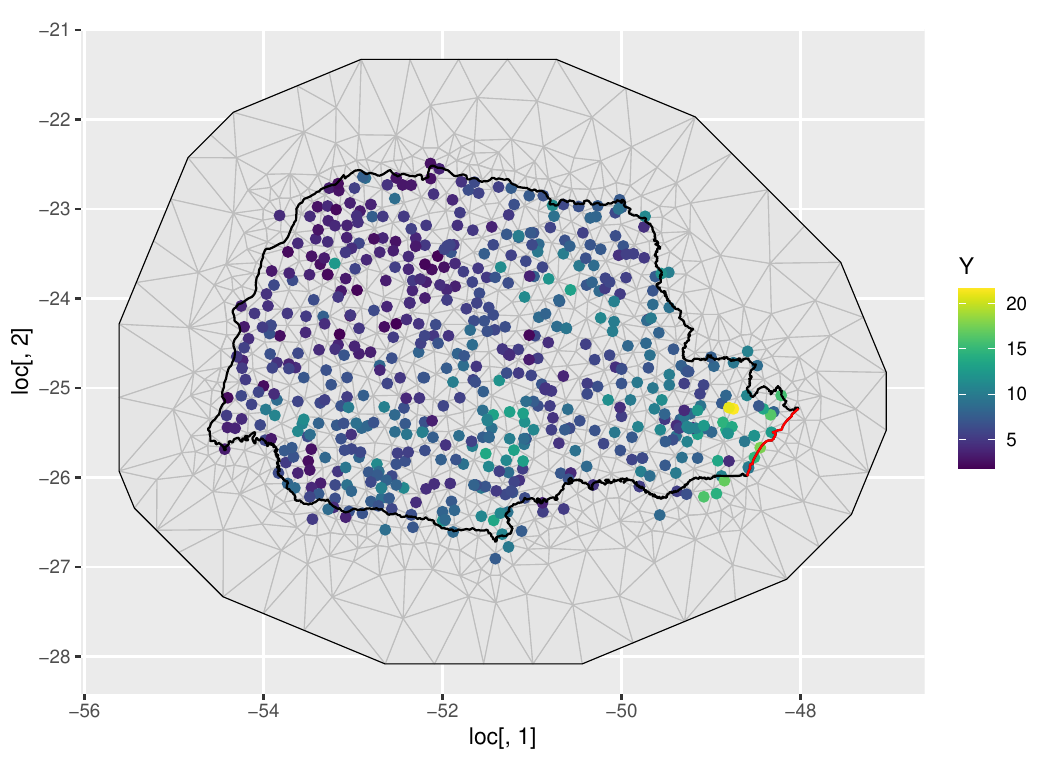}
\caption{\label{fig:inladata} Precipitation data, region boundary and mesh. The red line shows the coast.}
\end{figure}

Figure~\ref{fig:inladata} shows the data, where the red line shows the coast line, and we expect the distance to the coast to be a good covariate for precipitation. 
This covariate is not available, so let us calculate it for each observation location:

\begin{Schunk}
\begin{Sinput}
R> dcov <- apply(spDists(loc, PRborder[1034:1078, ], longlat = TRUE), 1, min)
\end{Sinput}
\end{Schunk}

As precipitation data are non-negative, we assume that the data is Gamma distributed, with mean $\mu$ and variance $\mu^2/\phi$, where $1/\phi$ is a dispersion parameter. The mean is modeled using a stochastic model that includes both the distance to the coast as a covariate and a Gaussian field, resulting in the latent Gaussian model for the precipitation measurements
\begin{align*} 
    y_i\mid \mu(s_i), \theta &\sim \Gamma(\mu(s_i),\phi)\\ 
    \log (\mu(s)) &= \eta(s) = I + f(c(s))+u(s)\\ 
    \theta &\sim \pi(\theta),
\end{align*}
where $y_i$ denotes the measurement taken at location $s_i$, $c(s)$ is the covariate whose effect is captured through a random walk model $f$, $I$ an intercept, and $u(s)$ is a mean-zero Gaussian Matérn field, and $\theta$ is a vector containing all parameters of the model.

We can use \pkg{fmesher} for creating the mesh. Let us create a mesh which is based 
on a non-convex hull to avoid adding many small triangles outside the domain 
of interest. 

\begin{Schunk}
\begin{Sinput}
R> dom <- fm_nonconvex_hull(loc, -0.03, -0.05, resolution = c(100, 100))
R> mesh <- fm_mesh_2d(boundary = dom, max.edge = c(0.3, 1), cutoff = 0.1)
\end{Sinput}
\end{Schunk}

The resulting mesh is shown in Figure~\ref{fig:inladata}.
We now create the observations matrix, that connects the mesh to the observation 
locations and then create the \pkg{rSPDE} model. For this task, as we mentioned earlier, we need to use the \pkg{rSPDE}-specific function, \code{rspde.make.A}.
The reason for the need of this specific function is that the size of
the matrix matrix depends on the order of the rational approximation, and 
whether or not we estimate the smoothness parameter or not. If we estimate the smoothness parameter and use the default order of 1, there is no need to specify anything except for the mesh and the observation locations
\begin{Schunk}
\begin{Sinput}
R> Abar <- rspde.make.A(mesh = mesh, loc = loc)
\end{Sinput}
\end{Schunk}

We now construct the model through the \code{rspde.matern} function. Since we are using the default order and will estimate the smoothness, all we need to supply is the mesh

\begin{Schunk}
\begin{Sinput}
R> model <- rspde.matern(mesh = mesh)
\end{Sinput}
\end{Schunk}

As the model is created, it sets default priors for the parameters, log-normal for $\kappa$ and $\tau$ and a $\beta$-distribution on $(0,2)$ for $\nu$ (as the upper bound for $\nu$ is set to the default value). We provide the details for these options, and other ways of adjusting the model specification in Appendix~\ref{sec:inladetails}. Non-stationary models can also be created using this function by specifying matrices \code{B.kappa} and \code{B.tau} in the same way as in \code{spde.matern.operators} mentioned above and as in \code{inla.spde2.matern} in \pkg{INLA}. 

As for the standard SPDE models in \pkg{INLA}, the final steps of model creation are to define the indices for the random field and to define the observation stack. The stack is created exactly in the same way as for standard SPDE models in \pkg{INLA}, but the indices need to be build using the \code{rspde.make.index} function. Again, since we are using the default order and are estimating the smoothness, we only need to provide the mesh and give a name to the field, otherwise the \code{rspde.order} and \code{nu} arguments are used in the same was as for the model creation.  

\begin{Schunk}
\begin{Sinput}
R> ind <- rspde.make.index(name = "u", mesh = mesh)
R> stk <- inla.stack(data = list(y = Y), A = list(Abar, 1), 
+                    effects = list(c(ind), list(sdist = inla.group(dcov),
+                                                Intercept = 1)))
\end{Sinput}
\end{Schunk}

Here the observation matrix $\boldsymbol{A}$ is applied to the spatial effect while an identity observation matrix, denoted by $1$, 
is applied to the covariates and the intercept. This means the covariates are unaffected 
by the observation matrix. 

The observation matrices in \code{A=list(Abar,1)} are used to link the corresponding elements in the effects-list to the observations. Thus in our model the latent spatial field is linked to the log-expectation of the observations, i.e. $\eta(s)$, through the matrix $\boldsymbol{A}$.
The covariate and the intercept, on the other hand, are linked directly to $\eta(s)$. 

We now specify the model using the random walk model for the covariate as follows:
\begin{Schunk}
\begin{Sinput}
R> fs <- y ~ -1 + Intercept + f(sdist, model = "rw1") + f(u, model = model)
\end{Sinput}
\end{Schunk}
Here \code{-1} is added to remove the implicit intercept, which is replaced 
by the \code{Intercept} term. 

To fit the model we proceed as in the standard SPDE approach and
we simply call \code{inla}. 

\begin{Schunk}
\begin{Sinput}
R> rspde_fit <- inla(fs, family = "Gamma", 
+                    data = inla.stack.data(stk), verbose = FALSE, 
+                    control.predictor = list(A = inla.stack.A(stk)))
\end{Sinput}
\end{Schunk}

We can look at some summaries of the posterior distributions for the parameters, 
for example the fixed effects (i.e. the intercept) and the hyper-parameters by writing 
\code{summary(rspde_fit)}. This provides the summary of the random field parameters 
in the internal parameterization used in the optimization. Since we in this case 
used a stationary model, we have three internal parameters $\theta_1, \theta_2, \theta_3$ which are linked to the parameters $\kappa, \tau, \nu$ of the SPDE model 
through $\tau = \exp(\theta_1)$, $\kappa = \exp(\theta_2)$ and 
$\nu = \nu_{UB}\Big(\frac{\exp(\theta_3)}{1+\exp(\theta_3)}\Big)$, 
where $\nu_{UB}$ is the value of the upper bound for the smoothness parameter $\nu$,
which in this case was set to the default value of $2$. 
We can obtain outputs with respect to parameters in the original scale by
using the function \code{rspde.result}:

\begin{Schunk}
\begin{Sinput}
R> result_fit <- rspde.result(rspde_fit, "u", model)
R> summary(result_fit)
\end{Sinput}
\begin{Soutput}
          mean       sd 0.025quant 0.5quant 0.975quant     mode
tau   1.036800 1.133690  0.1334490 0.683098    4.11357 0.326799
kappa 5.223220 2.087380  2.2045000 4.873980   10.29090 4.213210
nu    0.486354 0.308374  0.0739538 0.423521    1.22533 0.231801
\end{Soutput}
\end{Schunk}

This function is reminiscent to the \code{inla.spde.result} in \pkg{INLA}  with the main difference that it has \code{summary} and \code{plot} methods implemented.

\begin{figure}[t!]
\centering
\includegraphics{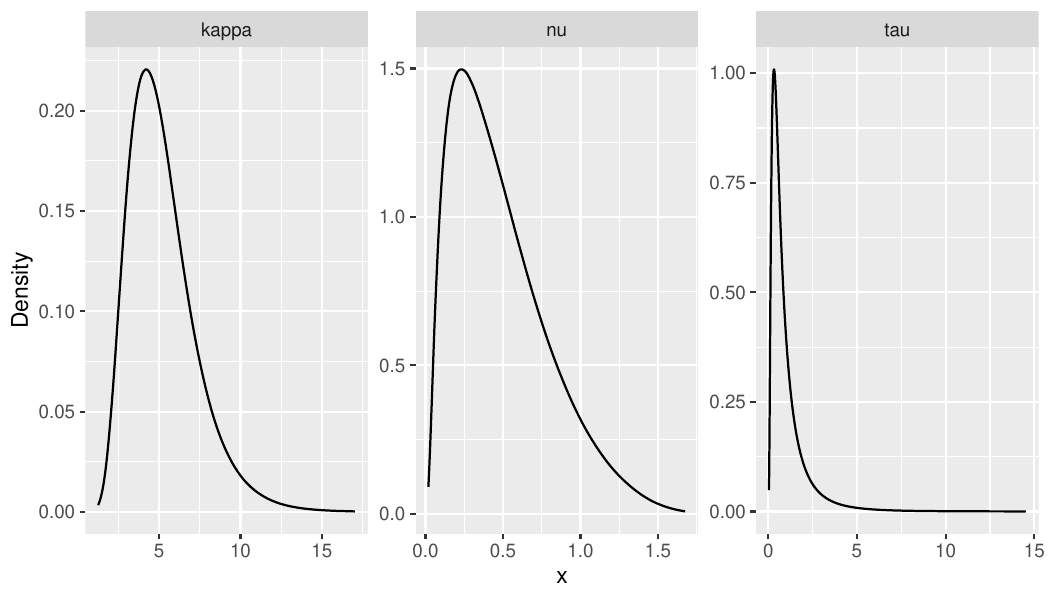}
\caption{Posterior distributions for the parameters.}
\label{fig:inlares}
\end{figure}

To create plots of the posterior marginal densities, we can use the \code{gg_df} function, which creates  data frames adapted for plotting using the \pkg{ggplot2} package \citep{ggplot2}. Figure~\ref{fig:inlares} shows the posterior marginal densities of the three parameters and is created as follows

\begin{Schunk}
\begin{Sinput}
R> posterior_df_fit <- gg_df(result_fit)
R> ggplot(posterior_df_fit) + geom_line(aes(x = x, y = y)) + 
+  facet_wrap(~parameter, scales = "free") + labs(y = "Density")
\end{Sinput}
\end{Schunk}

If we instead want the posteriors for the marginal standard deviation and practical correlation range, \code{rspde.result} can be called with the argument \code{parameterization = "matern"}. 

\section[inlabru interface]{\pkg{inlabru} interface}\label{sec:inlabru}

In this section, we illustrate the \pkg{inlabru} interface of \pkg{rSPDE} using the same data as in the previous section. The mesh and model creation for \pkg{inlabru} is exactly the same as for \pkg{inla}, so we use the previously defined data, mesh and \pkg{rSPDE} model. 

The difference with the \pkg{inlabru} interface is that we do not need to construct the observation matrix, indices or the stack. Instead, we simply build a data frame that contains the data and spatial coordinates.

\begin{Schunk}
\begin{Sinput}
R> prdata <- data.frame(long = loc[,1], lat = loc[,2], 
+                       sdist = inla.group(dcov), y = Y)
R> prdata <- st_as_sf(prdata, coords = c("long", "lat"), crs = 4326)
\end{Sinput}
\end{Schunk}

Having defined the \pkg{rSPDE} model \code{model} and the data frame, we can now directly define the linear predictor and fit the model using \code{bru}:
\begin{Schunk}
\begin{Sinput}
R> cmp <- y ~ Intercept(1) + distSea(sdist, model="rw1") + 
+      field(geometry, model = model)
R> fit <- bru(cmp, data = prdata, family = "Gamma")
\end{Sinput}
\end{Schunk}

\begin{figure}[t!]
\centering
\includegraphics{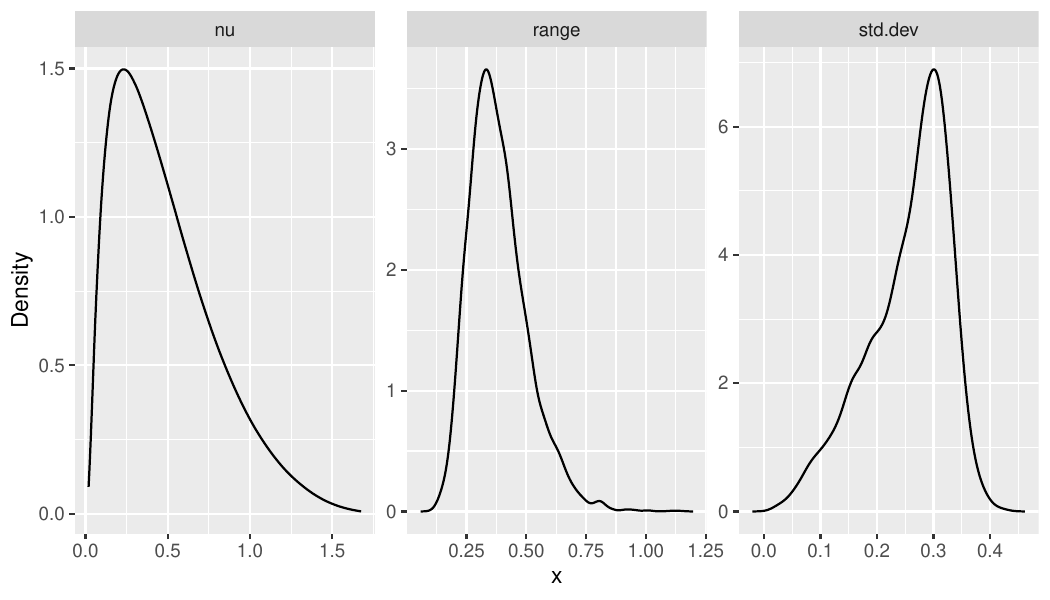}
\caption{\label{fig:brures} Posterior distributions for the parameters.}
\end{figure}

As for the \pkg{INLA} results, we can obtain the posterior distributions for the parameters using \code{rspde.result}, and plot the results using \code{gg_df} in combination with \pkg{ggplot2}. Let us illustrate this by plotting the posteriors for the marginal standard deviation and practical correlation range. The result is shown in Figure~\ref{fig:brures}

Let us now see how we can obtain predictions of the expected precipitation on 
a dense grid in the region. We begin by creating the grid in which we want to do the predictions. To this end, we can use the \code{fm_evaluator} function of the \pkg{fmesher} package:

\begin{Schunk}
\begin{Sinput}
R> grid <- fm_evaluator(mesh, xlim = range(PRborder[, 1]),
+                       ylim = range(PRborder[, 2]), dims = c(150, 100))
\end{Sinput}
\end{Schunk}

Let us remove the locations of the mesh that are outside the region of interest and create a data frame with the coordinates. Since we are using distance to the sea as a covariate, we also need to calculate this covariate for the prediction locations and then add it to the data frame. 

\begin{Schunk}
\begin{Sinput}
R> xy_in <- inout(grid$lattice$loc, PRborder)
R> loc_prd <- grid$lattice$loc[xy_in, ]
R> prd_df <- data.frame(x1 = loc_prd[,1], x2 = loc_prd[,2])
R> prd_df <- st_as_sf(prd_df, coords = c("x1", "x2"), crs = 4326)
R> seaDist_prd <- apply(spDists(loc_prd, PRborder[1034:1078, ],
+                               longlat = TRUE), 1, min)
R> prd_df$sdist <- seaDist_prd
\end{Sinput}
\end{Schunk}

We can now compute the prediction and plot the predicted mean as 
\begin{Schunk}
\begin{Sinput}
R> pred <- predict(fit, prd_df, ~exp(Intercept + field + distSea))
R> ggplot() + gg(pred, geom = "tile", aes(fill = mean)) +
+      geom_raster() + scale_fill_viridis()
\end{Sinput}
\end{Schunk}

\begin{figure}[t!]
\centering
\includegraphics{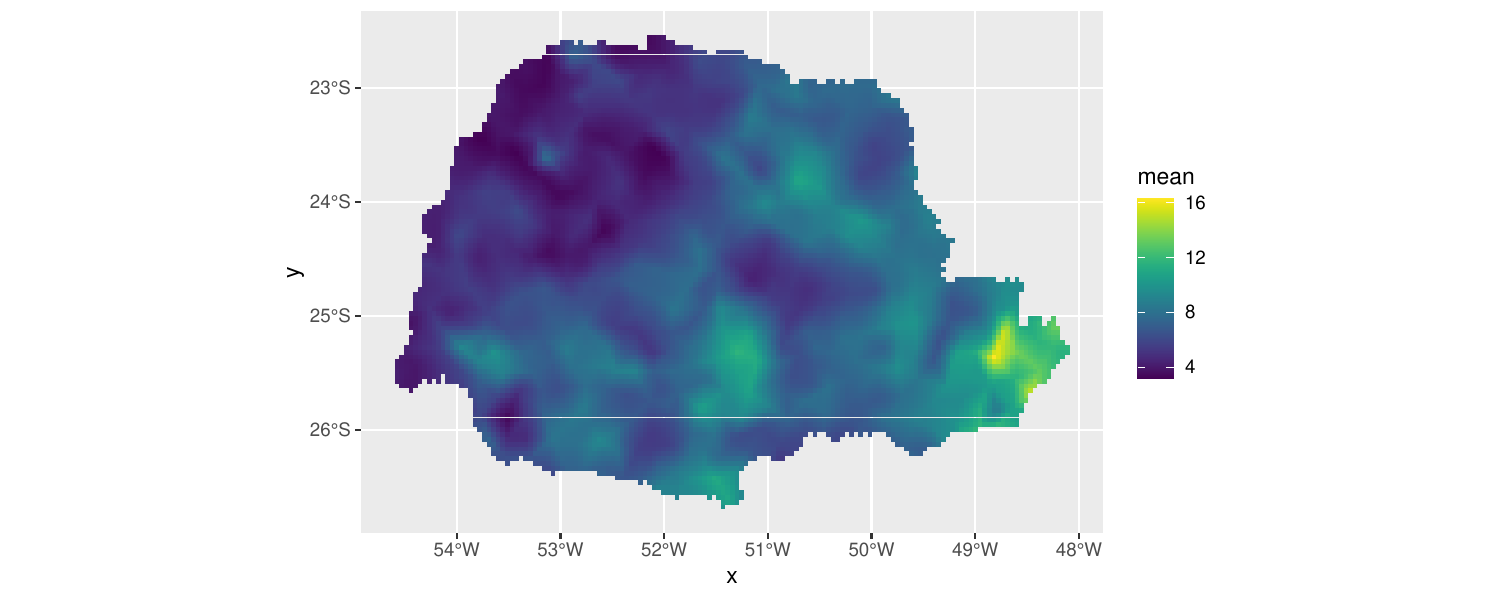}
\caption{Posterior mean of $\mu(s)$.}
\label{fig:brupred}
\end{figure}

The result is shown in Figure~\ref{fig:brupred}.

\newpage
\section[MetricGraph interface]{\pkg{MetricGraph} interface} \label{sec:metricgraph}

The \pkg{rSPDE} package also has an interface to the \pkg{MetricGraph} package,
which means that one can use the two packages to define Whittle--Mat\'ern fields
with general smoothness on compact metric graphs, as introduced in \cite{lenin}. 

To illustrate this, we begin by loading the package and then defining a simple 
metric graph from the logo of the \pkg{MetricGraph} package
\begin{Schunk}
\begin{Sinput}
R>   library(MetricGraph)
R>   graph <- metric_graph$new()
\end{Sinput}
\end{Schunk}

\begin{figure}[t!]
\centering
\includegraphics{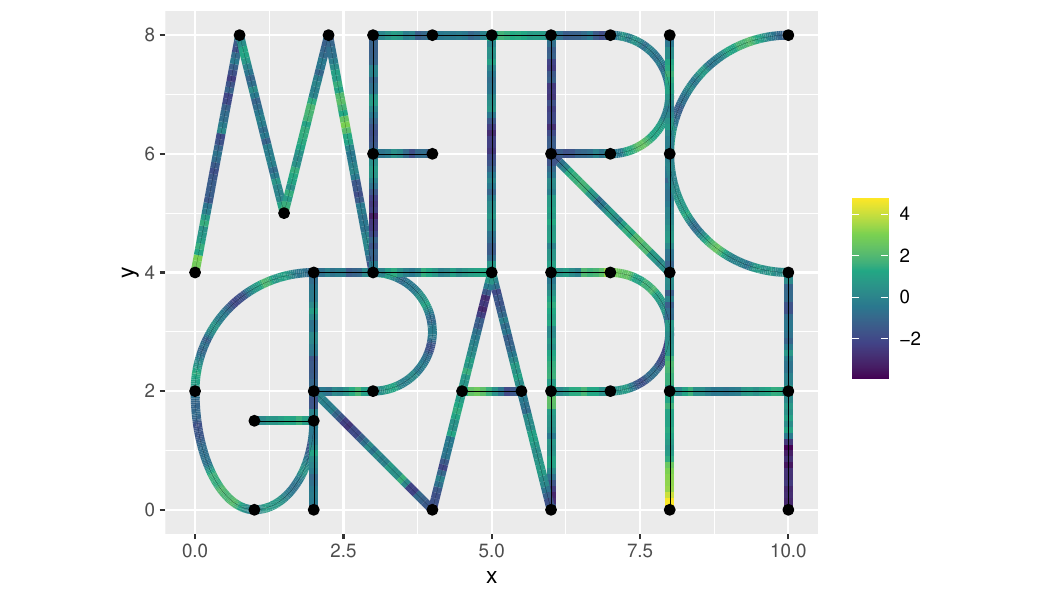}
\caption{\label{fig:graph2} A realization of a Whittle--Mat\'ern field on a compact metric graph.}
\end{figure}

The graph is shown in Figure~\ref{fig:graph2}. 
To construct a FEM approximation of a Whittle--Matérn field with general smoothness, 
we must first construct a mesh on the graph. 
\begin{Schunk}
\begin{Sinput}
R>   graph$build_mesh(h = 0.1)
\end{Sinput}
\end{Schunk}

In the command \code{build_mesh}, the argument \code{h} decides the largest spacing between nodes in
the mesh. The mesh can be visualized by \code{graph$plot(mesh=TRUE)}

We are now ready to specify the model \eqref{eq:spde} 
for the Whittle--Matérn field $u$ on this graph. 
For this, we use the \code{matern.operators} function which also accepts metric graphs 
as inputs.
\begin{Schunk}
\begin{Sinput}
R>   par <- list(sigma = 1.3, r = 1, nu = 0.8)
R>   op <- matern.operators(nu = par$nu, range = par$r, sigma = par$sigma, 
+                           parameterization = "matern", graph = graph)                     
\end{Sinput}
\end{Schunk}
As can be seen in the code, we here used the Mat\'ern parameterization so that the practical 
correlation range and marginal standard deviation is specified. 

Let us simulate the field $u$ at the mesh locations and plot the result:
\begin{Schunk}
\begin{Sinput}
R> u <- simulate(op)
R> graph$plot_function(X = u, vertex_size = 2, edge_width = 2)
\end{Sinput}
\end{Schunk}

Let us now generate some observation locations and construct the corresponding 
observation matrix. This can be done
by the function \code{fem_basis} in the metric graph object. 
\begin{Schunk}
\begin{Sinput}
R> obs_per_edge <- 10
R> loc <- NULL
R> for(i in 1:graph$nE) {
+    loc <- rbind(loc, cbind(rep(i,obs_per_edge), runif(obs_per_edge)))
+  }
R> n_obs <- obs_per_edge*graph$nE
R> A <- graph$fem_basis(loc)
\end{Sinput}
\end{Schunk}

We can now use the \code{simulate} function to sample the process $10$ times 
and generate observed values of the process under Gaussian measurement noise.
\begin{Schunk}
\begin{Sinput}
R>   n_rep <- 10
R>   u_rep <- simulate(op, nsim = n_rep)
R>   Y <- A %*% u_rep + 0.3 * matrix(rnorm(n_obs * n_rep), ncol = n_rep)
\end{Sinput}
\end{Schunk}

We can now add the data with replicates to the graph:

\begin{Schunk}
\begin{Sinput}
R>   df <- data.frame(y=as.vector(Y), 
+                     edge_number = rep(loc[,1], n_rep),
+                     distance_on_edge = rep(loc[,2], n_rep),
+                     repl = rep(1:n_rep, each = n_obs))
R> graph$add_observations(data = df, group = "repl", normalized = TRUE)
\end{Sinput}
\end{Schunk}

Having generated some data on the graph, we can now estimate the model based on this data using either
the \code{graph_lme} function or using the \pkg{INLA} or \pkg{inlabru} interfaces. As an illustration,
let us estimate the model using \pkg{inlabru}.

We start by creating the \pkg{rSPDE} model using the \code{rspde.metric_graph} function:

\begin{Schunk}
\begin{Sinput}
R>   model <- rspde.metric_graph(graph)
\end{Sinput}
\end{Schunk}

To use \pkg{bru}, we must have the data in a data frame. We can extract the data from the 
graph in the correct format by using the \code{graph_data_spde} function. To indicate 
that we want all replicates, we specify the \code{repl} argument
\begin{Schunk}
\begin{Sinput}
R>   data_rspde <- graph_data_rspde(model, repl = ".all", repl_col = "repl")
\end{Sinput}
\end{Schunk}

We can now define the \pkg{bru} component formula, passing the \code{repl} as the \code{replicate} argument:
\begin{Schunk}
\begin{Sinput}
R>   cmp_rep <- y ~ -1 + field(cbind(.edge_number, .distance_on_edge), 
+                              model = model, replicate = repl)
\end{Sinput}
\end{Schunk}
Now, we are ready to fit the model using \code{bru}:
\begin{Schunk}
\begin{Sinput}
R>   bru_fit <- bru(cmp_rep, data=data_rspde[["data"]])
\end{Sinput}
\end{Schunk}

We can use the \code{rspde.result} function as before to obtain summaries 
of the parameters. Let us extract these and compare to the true parameters:
\begin{Schunk}
\begin{Sinput}
R>   res <- rspde.result(bru_fit, "field", model)
R>   print(data.frame(parameter = c("std.dev", "range", "nu"),
+                     true = c(par$sigma, par$r, par$nu),
+                     mean = c(res$summary.std.dev$mean, 
+                              res$summary.range$mean,
+                              res$summary.nu$mean),
+                     mode = c(res$summary.std.dev$mode,
+                              res$summary.range$mode,
+                              res$summary.nu$mode)))
\end{Sinput}
\begin{Soutput}
  parameter true      mean      mode
1   std.dev  1.3 1.3288275 1.3291485
2     range  1.0 0.9682099 0.9716748
3        nu  0.8 0.9525355 0.9301602
\end{Soutput}
\end{Schunk}

We could also plot the posterior marginal densities of the parameters with the help 
of the \code{gg_df} function as before. 

Let us finally  do prediction for the 10th replicate. 
We start by building the data list with the prediction locations:

\begin{Schunk}
\begin{Sinput}
R>   data_prd_list <- graph$get_mesh_locations(bru = TRUE)
R>   data_prd_list[["repl"]] <- rep(10, nrow(data_prd_list))
\end{Sinput}
\end{Schunk}

We then obtain predictions for this replicate and plot the results:

\begin{Schunk}
\begin{Sinput}
R>   y_pred <- predict(bru_fit, newdata=data_prd_list, 
+                      ~field_eval(cbind(.edge_number, .distance_on_edge), 
+                                  replicate = repl))
R>   y_pred <- process_rspde_predictions(y_pred, graph = graph, 
+                                        PtE = data_prd_list)
R>   plot(y_pred, edge_width = 2, vertex_size = 2)
\end{Sinput}
\end{Schunk}

\begin{figure}[t!]
\centering
\includegraphics{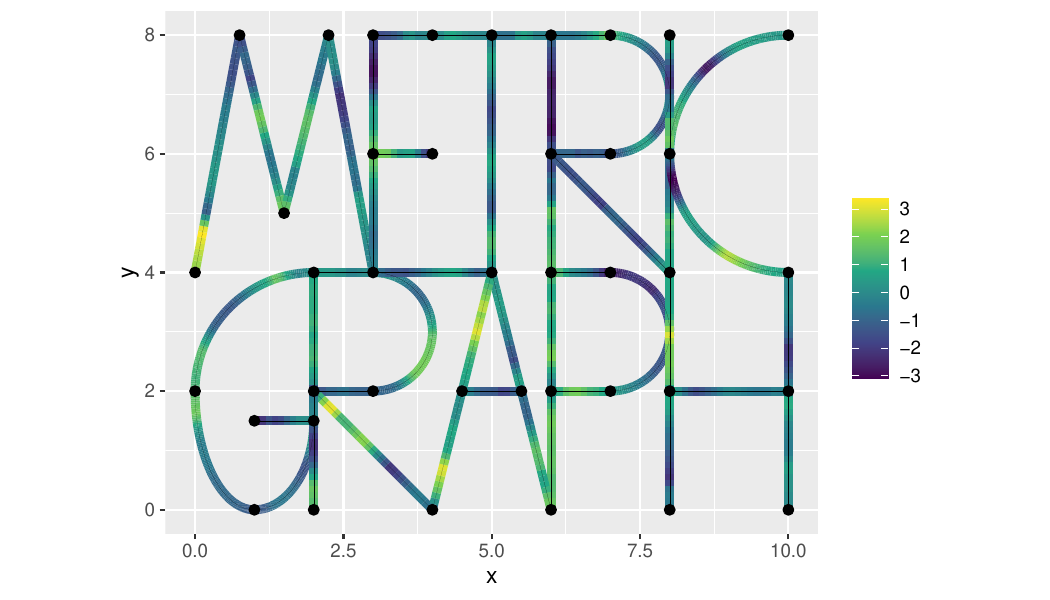}
\caption{\label{fig:graphpred} Predictions on the graph.}
\end{figure}

The predictions are shown in Figure~\ref{fig:graphpred}.

\section{Other SPDE-based models}\label{sec:othermodels}
Besides the (generalized) Whittle--Mat\'ern fields, \pkg{rSPDE} contains a few other models which we will briefly introduce in this section. 

\subsection{Anisotropic Whittle--Mat\'ern fields}
For domains $D\subset \mathbb{R}^2$, the \pkg{rSPDE} package implements the anisotropic Matérn model
$$
    (I - \nabla\cdot (H\nabla))^{(\nu + 1)/2} u = c\sigma W, \quad \text{on }  D
$$
Where $H$ is a $2\times 2$ positive definite matrix, $\sigma, \nu >0$ and $c$ is a constant chosen such that $u$ would have the covariance function 
$$
r(h) = \frac{\sigma^2}{2^{\nu-1}\Gamma(\nu)} (\sqrt{h^T H^{-1} h})^{\nu} K_{\nu}(\sqrt{h^T H^{-1} h}),
$$
if the domain was $D = \mathbb{R}^2$, i.e., a stationary and anisotropic Matérn covariance function. 
The matrix $H$ is defined as 
$$
H = \begin{bmatrix}
h_x^2 & h_xh_y h_{xy}\\
h_xh_y h_{xy} & h_y^2
\end{bmatrix},
$$
with $h_x,h_y>0$ and $h_{xy} \in (-1,1)$. Non-fractional models of this type, and corresponding non-stationary versions were introduced by \citet{fuglstad2015} and fractional versions were investigated in \citet{hildeman2021deformed}. Thus, \pkg{rSPDE} currently implements a stationary version of the models considered in \citet{hildeman2021deformed}.

To define the model, the \code{matern2d.operators} function can be used. 
\begin{Schunk}
\begin{Sinput}
R> bnd <- fm_segm(rbind(c(0, 0), c(2, 0), c(2, 2), c(0, 2)), is.bnd = TRUE)
R> mesh_2d <- fm_mesh_2d(boundary = bnd, cutoff = 0.02, max.edge = c(0.05))
R> op <- matern2d.operators(hx = 0.1, hy = 0.1, hxy = 0.5, nu = 0.75, 
+                           sigma = 1, mesh = mesh_2d)
\end{Sinput}
\end{Schunk}

\begin{figure}[t!]
\centering
\includegraphics{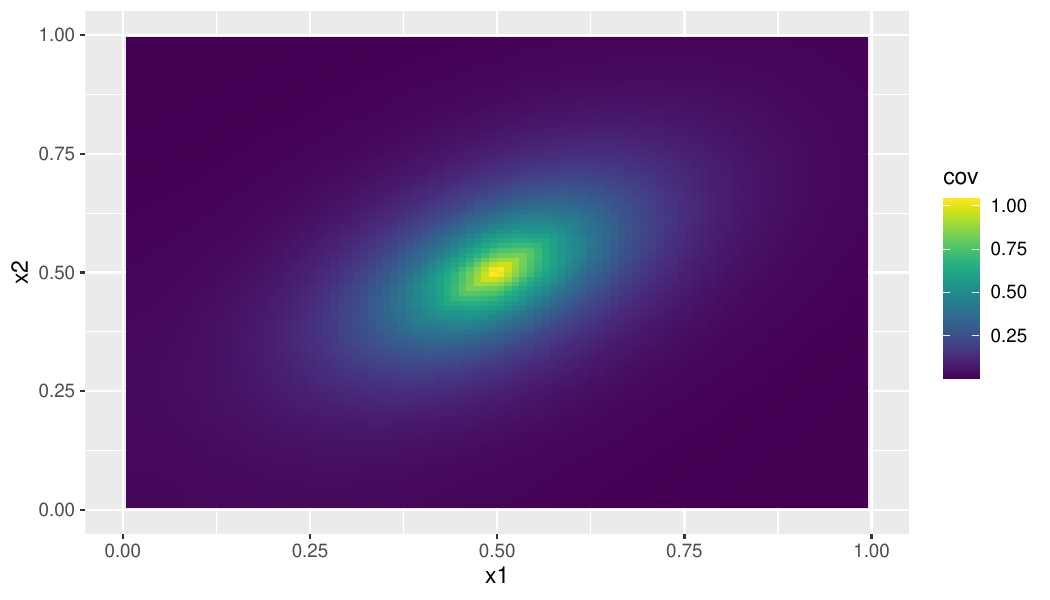}
\caption{\label{fig:aniso} Anisotropic Mat\'ern covariance.}
\end{figure}

The \code{matern2d.operators} object has an \code{cov_function_mesh} method which can be used 
evaluate the covariance function on the mesh. For example
\begin{Schunk}
\begin{Sinput}
R> r <- op$cov_function_mesh(matrix(c(0.5,0.5),1,2))
R> proj <- fm_evaluator(mesh_2d, dims = c(100, 100), 
+                       xlim = c(0,1), ylim = c(0,1))
R> r_mesh <- fm_evaluate(proj, field = as.vector(r))
R> cov_df <- data.frame(x1 = proj$lattice$loc[,1],
+                       x2 = proj$lattice$loc[,2], 
+                       cov = c(r_mesh))
R> ggplot(cov_df, aes(x = x1, y = x2, fill = cov)) + geom_raster() + 
+      xlim(0,1) + ylim(0,1) + scale_fill_viridis()
\end{Sinput}
\end{Schunk}
The result can be seen in Figure~\ref{fig:aniso}.

Simulation and kriging based on this method can be done using the \code{simulate} and \code{predict} methods as for any other model implemented in \pkg{rSPDE}. The model can also be fitted to data using \code{rspde_lme}, and predictions based on the fitted model 
can be obtained using \code{predict} on the fitted object. 

To include the model in Bayesian models which can be fitted to data using \pkg{INLA} or \pkg{inlabru}, the model is instead defined using the \code{rspde.anisotropic2d} function:
\begin{Schunk}
\begin{Sinput}
R> model_aniso <- rspde.anistropic2d(mesh = mesh_2d)
\end{Sinput}
\end{Schunk}
Once the model has been defined, it can be used in the same way as any other \pkg{rSPDE} model in combination with  \pkg{INLA} or \pkg{inlabru}. We refer to \url{https://davidbolin.github.io/rSPDE/articles/anisotropic.html} for further details on these models. 

Future work includes implementing the non-stationary versions of these models.

\subsection{Intrinsic random fields}
Intrinsic random fields are used in several areas of research. An example of an intrinsic random field is the solution to 
$$
(-\Delta)^{\beta/2}(\tau u) = \mathcal{W},
$$ 
where $\beta > d/2$ and $d$ is the dimension of the spatial domain. Thus, $u$ here can be viewed as a Whittle--Mat\'ern field with $\kappa = 0$. Suppose that the equation is posed on a compact subset $D\subset \mathbb{R}^2$, and that the Laplacian is equipped with Neumann boundary conditions. The solution $u$ is then intrinsic in the sense that the field is invariant to the addition of a constant, which can be a useful property if used as priors in Bayesian models. 

If we consider these models on the space of functions on $D$ which are orthogonal to the constants, the fields are then proper Gaussian fields, which can be implemented similarly to the standard Whittle--Mat\'ern fields. The fields are implemented using the \code{intrinsic.operators} function in \pkg{rSPDE}:
\begin{Schunk}
\begin{Sinput}
R> mesh_2d <- fm_mesh_2d(boundary = bnd, cutoff = 0.04, max.edge = c(0.09))
R> op <- intrinsic.operators(tau = 0.2, beta = 1.8, mesh = mesh_2d, m = 4)
\end{Sinput}
\end{Schunk}
To verify that the \pkg{rSPDE} model is approximating the true model, we compare the variogram
of the approximation with the true variogram. 
The variogram can be computed using the \code{variogram} method of the operator object.
We consider the variogram function $\gamma(\boldsymbol{s}_0,\boldsymbol{s})$ for a fixed location $\boldsymbol{s}_0 = (1,1)$, and look at this function for all mesh locations. We then sort these with respect to the distance to $\boldsymbol{s}_0$ and compare with the true variogram  which is implemented in \code{variogram.intrinsic.spde}.
\begin{Schunk}
\begin{Sinput}
R> s0 <- matrix(c(1,1),1,2)
R> Gamma <- op$variogram(s0)
R> vario <- variogram.intrinsic.spde(s0, mesh_2d$loc[,1:2], tau = 0.2,
+                                    beta = 1.8, L = 2, d = 2)
R> d = sqrt((mesh_2d$loc[,1]-s0[1])^2 +  (mesh_2d$loc[,2]-s0[2])^2)
R> plot(d, Gamma, xlim = c(0,0.7), ylim = c(0,4))
R> lines(sort(d),sort(vario),col=2, lwd = 2)
\end{Sinput}
\end{Schunk}

\begin{figure}[t!]
\centering
\includegraphics{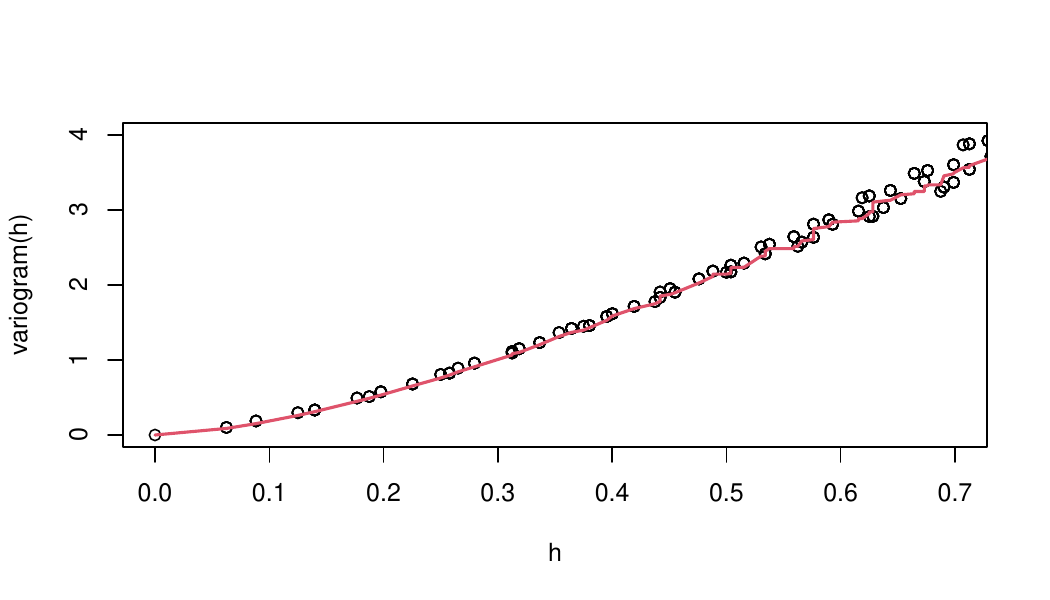}
\caption{\label{fig:intrinsic} Variogram of the \pkg{rSPDE} approximation of an intrinsic field (black) and the corresponding variogram of the exact model (red).}
\end{figure}

Simulation and prediction for intrinsic fields can be done using the \code{simulate} and \code{predict} functions, respectively. By default, the field is simulated with a zero-integral constraint to handle the intrinsic nature of the field. Estimation can be done using \code{rspde_lme} or through \pkg{INLA}. 
The \pkg{INLA} version of the model is implemented as 
\begin{Schunk}
\begin{Sinput}
R> rspde_model <- rspde.intrinsic(mesh = mesh_2d)
\end{Sinput}
\end{Schunk}
and the model can then be treated as any other SPDE-based model in \pkg{INLA} or \pkg{inlabru}. 
In particular, both $\tau$ and $\beta$ can be estimated from data. 
We refer to \url{https://davidbolin.github.io/rSPDE/articles/intrinsic.html} for further details on these models.

The \pkg{rSPDE} package also implements a more general class of intrinsic models, 
defined as  
$$
(-\Delta)^{\beta/2}(\kappa^2-\Delta)^{\alpha/2}(\tau u) = \mathcal{W},
$$ 
where $\alpha + \beta > d/2$ and $d$ is the dimension of the spatial domain. The models can be specified through the \code{intrinsic.matern.operators} and simulated from using the corresponding \code{simulate} method. Currently,
these models are fully implemented in \pkg{rSPDE}, where prediction can be done through the \code{predict} method and estimation of all model parameters (including the two smoothness parameters) can be done through \code{rspde_lme}. 

The more general models currently have a partial \pkg{INLA} implementation. Specifically, the model can be used in  \pkg{INLA} if $\beta$ is fixed at $1$ and $\alpha$ is fixed at either $1$ or $2$.
Future work includes implementing the fractional versions of these models for \pkg{INLA}, and the theory for the models will be introduced in a future paper.  

\subsection{Spatio-temporal models with drift}
The \pkg{rSPDE} package implements the following spatio-temporal model 
$$
d u + \gamma(\kappa^2 + \kappa^{d/2}\rho\cdot \nabla - \Delta)^{\alpha} u = dW_Q, \quad \text{on } T\times D,
$$
where $T$ is a temporal interval and $D$ is a spatial domain which can be an interval or a bounded subset of $\mathbb{R}^2$. Here $\kappa>0$ is a spatial range parameter, $\rho$ is a drift parameter which is in $\mathbb{R}$ for spatial domains that are intervals or metric graphs, and in $\mathbb{R}^2$ for spatial domains which are bounded subsets of $\mathbb{R}^2$. Further, $W_Q$ is a $Q$-Wiener process with spatial covariance operator $\sigma^2(\kappa^2 - \Delta)^{-\beta}$, where $\sigma^2$ is a variance parameter. Thus, the model has two smoothness parameters $\alpha$ and $\beta$ which are assumed to be integers. The model is therefore a generalization of the spatio-temporal models introduced in \citet{lindgren2024spacetime}, where the generalization is to allow for drift. The model can also be viewed as an alternative to the spatio-temporal models in \citet{Clarotto2024} where a slightly different spatio-temporal model is constructed, which is discretized using a different strategy. 

The model is implemented using a FEM discretization of the corresponding precision operator
$$
\sigma^{-2}(d + \gamma(\kappa^2 + \kappa^{d/2}\rho\cdot \nabla - \Delta)^{\alpha})(\kappa^2 - \Delta)^{\beta} (d + \gamma(\kappa^2 - \kappa^{d/2}\rho\cdot \nabla - \Delta)^{\alpha}),
$$
in both space and time, similarly to the discretization introduced in \citet{lindgren2024spacetime}. This parameterization of the drift term, using $\rho \kappa^{d/2}$ is chosen to simplify the enforcement of theoretical bounds on the range of $\rho$, ensuring that the equation remains well-posed.

The function \code{spacetime.operators} is used to define the model. The function requires specifying the two smoothness parameters, and the discretization points for the spatial and temporal discretizations. The spatial discretization can be specified through a mesh object from the \pkg{fmesher} package or as the mesh nodes for models on intervals. The temporal discretization can be specified either through the mesh nodes or by providing a mesh object. 

Assume that we want to define a model on the spatial interval $[0,20]$ and the temporal domain $[0,10]$. We can then simply specify the mesh nodes and define the model as 
\begin{Schunk}
\begin{Sinput}
R> s <- seq(from = 0, to = 20, length.out = 101)
R> t <- seq(from = 0, to = 10, length.out = 21)
R> op <- spacetime.operators(space_loc = s, time_loc = t,
+                            kappa = 5, sigma = 10, alpha = 1,
+                            beta = 1, rho = 1, gamma = 1/20)
\end{Sinput}
\end{Schunk}

The \code{spacetime.operators} object has a \code{plot_covariances} method which for univariate spatial domains simply plots the covariance $C(u(s,t), u(s_0, t_0))$ for a fixed spatio-temporal location $(s_0, t_0)$ specified by the indices in the spatial and temporal discretizations. 

\begin{figure}[t!]
\centering
\includegraphics{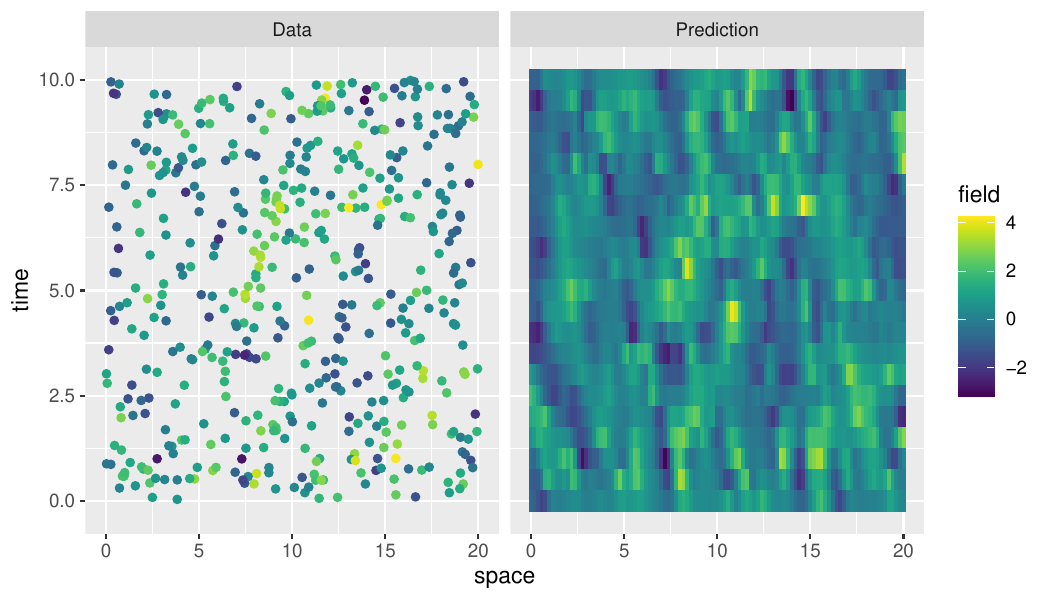}
\caption{\label{fig:st} Spatio-temporal data and kriging prediction.}
\end{figure}

We can simulate from the model using \code{simulate} and there is built-in support for kriging prediction using \code{predict}. The \code{predict} method requires specifying the projection matrices which links the mesh nodes to the observation and prediction locations, which can be obtained using the \code{make_A} method of the \code{spacetime.operators} object. For example, in the following code, we generate some data and compute the prediction. The results are shown in Figure~\ref{fig:st}.
\begin{Schunk}
\begin{Sinput}
R> u <- simulate(op)
R> loc <- data.frame(x = max(s)*runif(500), t = max(t)*runif(500))
R> A <- op$make_A(loc$x, loc$t)
R> Y <- as.vector(A%*%u + 0.01*rnorm(500))
R> Aprd <- op$make_A(rep(s, length(t)), rep(t, each = length(s)))
R> u_krig <- predict(op, A = A, Aprd = Aprd, Y = Y, sigma.e = 0.01)
\end{Sinput}
\end{Schunk}

To estimate the model parameters based on this data, we can use \code{rspde_lme} or \code{inlabru}. For this, we collect the data in a data frame, that also contains the observation locations.

\begin{Schunk}
\begin{Sinput}
R> df <- data.frame(y = as.matrix(Y), space = loc$x, time = loc$t)
R> res <- rspde_lme(y ~ 1, loc = "space", loc_time = "time", 
+                   data = df, model = op)
\end{Sinput}
\end{Schunk}

Here, \code{y~1} indicates that we want to estimate a mean value of the model. The arguments \code{loc} and \code{loc_time} provide the names of the spatial and temporal coordinates in the data frame. As for other models fitted using \code{rspde_lme}, prediction can be done based on the fitted object. 
\begin{Schunk}
\begin{Sinput}
R> pred_data <- data.frame(x = rep(s,length(t)), t = rep(t,each=length(s)))
R> pred <- predict(res, newdata = pred_data, loc = "x", time = "t")
\end{Sinput}
\end{Schunk}

To fit the model using \code{inlabru}, we create a model object with the \code{rspde.spacetime}, and create the formula which requires the user to pass the index as a list containing the elements \code{space} with the spatial indices and \code{time} with the temporal indices. 
\begin{Schunk}
\begin{Sinput}
R> model <- rspde.spacetime(space_loc = s, time_loc = t, alpha = 1, beta=1)
R> cmp <- y ~ -1 + Intercept(1) + 
+      field(list(space=space, time = time), model = model)
R> bru_fit <- bru(cmp, data = df)
\end{Sinput}
\end{Schunk}
Currently, $\alpha$ and $\beta$ have to be fixed at integer values, and future work includes implementing the fractional versions of these models.

\section{Summary and discussion} \label{sec:discussion}
We have illustrated how to use \pkg{rSPDE} to define and work with fractional-order generalized Whittle--Mat\'ern fields. The \code{rspde.matern} function which defines these models for \pkg{INLA} and \pkg{inlabru} serves as a complete replacement of the \code{inla.spde2.matern} function, which creates the corresponding non-fractional models in \pkg{INLA}. The advantage in using models defined through  \code{rspde.matern} is that they allow for using models with a arbitrary fixed smoothness, or to keep the smoothness parameter as an unknown parameter that is estimated from data jointly with the other model parameters. Contrary to the SPDE models in \pkg{INLA}, \pkg{rSPDE} also allows for the creation of SPDE models on metric graphs, through the \code{rspde.metric_graph} function. On metric graphs, intervals and circles, \pkg{rSPDE} also allows for the creation of models that do not require a FEM discretization. These models are exact Markov representations for the case $\alpha\in\mathbb{N}$, and highly accurate approximations (where the accuracy can be controlled through the order of the rational approximation) when $\alpha$ is not an integer. 

Future work includes the addition of more models in the package and in particular extending the functionality for the models introduced in Section~\ref{sec:othermodels}.

\section*{Computational details}
The results presented in this paper were obtained using \proglang{R}~4.4.2 with the
\pkg{rSPDE}~2.5.0 package. \proglang{R} itself
and all packages used except \pkg{INLA} are available from the Comprehensive
\proglang{R} Archive Network (CRAN) at
\url{https://CRAN.R-project.org/}. 
\pkg{INLA} can be downloaded from \url{https://www.r-inla.org/download-install}.

\pkg{rSPDE} depends on the \pkg{Matrix} \citep{Matrix} package and imports \pkg{stats}, \pkg{methods}, \pkg{fmesher}, \pkg{lifecycle} \citep{lifecycle}, and \pkg{broom} \citep{broom}. To replicate the codes in this paper, the following packages are required: \pkg{INLA}, \pkg{ggplot2}, \pkg{splancs} \citep{splancs}, \pkg{rgl}, \pkg{optimParallel}, \pkg{inlabru}, \pkg{gridExtra} \citep{gridExtra}, \pkg{MetricGraph}, and \pkg{sf} \citep{sf}.

The latent models in \pkg{INLA} were implemented using \pkg{INLA}'s Cgeneric interface by writing the code in \proglang{C} \citep{c17}. For some models, we also utilized functions written in \proglang{C++} \citep{cpp17} in combination with the Eigen library \citep{eigenweb}.
\section*{Acknowledgments}
The authors thank Håvard Rue for the help with implementing the Cgeneric inteface in \pkg{INLA}, which the \pkg{INLA} interface is based on, and for setting up the structure for including the binary files of \pkg{rSPDE} in the \pkg{INLA} release. We also thank Finn Lindgren for the help with \pkg{inlabru} related issues and for implementing the mapper system in \pkg{inlabru} which made it possible to add support for \pkg{rSPDE} models in \pkg{inlabru}.

\begin{appendices}

\section{Code for the comparison}\label{app:comparison}
The following code generates the results for Table~\ref{tab:sim_res}.
\begin{Schunk}
\begin{Sinput}
R> source("assessment.R")
R> load("AllSimulatedTemps.RData")
R> sim_data <- st_as_sf(all.sim.data, coords = c("Lon", "Lat"), crs = 4326)
R> ok_obs<- !is.na(sim_data$MaskTemp) 
R> ok_real <- !is.na(sim_data$TrueTemp)
R> mesh <- inla.mesh.create(loc=cbind(sim_data$Lon, sim_data$Lat), 
+                           extend=list(offset=-0.5),
+                           refine=list(min.angle=30))
R> scores <- list()
R> for(m in 1:3) {
+      spde <- rspde.matern(mesh, nu=0.5, rspde.order = m)
+  
+      res <- bru(MaskTemp ~ Intercept(1) + field(coordinates, model=spde), 
+                 data = sim_data, family = "normal",
+                 options = list(control.inla = list(int.strategy = "eb"),
+                                verbose = FALSE))
+      
+      pred <- pred.obj(res$summary.linear.predictor[1:150000,"mean"], 
+                       sqrt(res$summary.linear.predictor[1:150000,"sd"]^2
+                            + 1/res$summary.hyperpar[1,"0.5quant"]))
+      
+      scores[[m]] <- calc.scores(pred[!ok_obs & ok_real,], 
+                                 sim_data$TrueTemp[!ok_obs & ok_real], 
+                                 m)
+      if(m==1){
+          scores.total <- scores[[1]]
+      } else {
+          scores.total <- cbind(scores.total, scores[[m]])
+      }
+  }
R> print(scores.total)
\end{Sinput}
\end{Schunk}
\section[Details on the INLA model specification]{Details on the \pkg{INLA} model specification}\label{sec:inladetails}

In this section, we discuss some of the options available when creating an \pkg{rSPDE} model. 

\subsection{Changing the priors}\label{app:priors}

Recal that the fitted \pkg{rSPDE} model in \pkg{INLA} contains the 
parameters $\theta_1, \theta_2, \theta_2$. 
In the default parameterization, these are linked to $\kappa$ and $\tau$ through
$\tau = \exp(\theta_1)$ and $\kappa = \exp(\theta_2)$. Alternatively, one set 
\code{parameterization = "matern"} to instead have  
$\sigma = \exp(\theta_1)$ and $\rho = \exp(\theta_2)$, where 
$\rho$ is the practical correlation range and $\sigma$ is the marginal standard deviation.

Priors can be set on either $(\kappa,\tau)$, which we refer to as the SPDE parameterization or on 
$(\rho,\sigma)$. By default, \code{rspde.matern} assumes a lognormal prior distribution for $\tau$ and $\kappa$, i.e. that $\theta_1$ and $\theta_2$ follow normal distributions. 
By default $\theta_1\sim N(\log(\tau_0), 10)$ and $\theta_2\sim N(\log(\kappa_0), 10)$ are independent. Here, $\kappa_0$ is suitably defined in terms of the mesh and $\tau_0$ is defined in terms of $\kappa_0$ and on the prior of the smoothness parameter. The parameters of these normal distributions can be changed via the \code{prior.tau} and \code{prior.kappa} arguments.

One can alternatively set priors for $(\rho, \sigma)$ in the Mat\'ern parameterization. 
We have two options for the prior in this case. By default, which is the option 
\code{prior.theta.param = "theta"},  \code{rspde.matern}  assumes a lognormal prior  
for $\sigma$ and $\rho$. This prior distribution is obtained by assuming
that $\theta_1\sim N(\log(\sigma_0), 10)$ and $\theta_2\sim N(\log(\rho_0), 10)$
are independent.
Here, $\rho_0$ is suitably defined in terms of the mesh and $\sigma_0$ is defined in terms of $\rho_0$ and on the prior of the smoothness parameter.
The parameters for the priors of $\theta_1$ and $\theta_2$ can be changed through the 
\code{prior.range} and \code{prior.std.dev} arguments. 

Another option is to set \code{prior.theta.param = "spde"}. In this case, a change of variables is performed.
So, we assume a lognormal prior on $\tau$ and $\kappa$. Then, by the relations 
between $\rho, \sigma, \nu$ and $\kappa, \tau, \nu$ we obtain a prior for $\rho$ and $\sigma$.

Finally, let us consider the smoothness parameter $\nu$. 
By default, $\nu$ is assumed to follow a beta distribution on the interval $(0,\nu_{UB})$, where $\nu_{UB}$ is the upper bound for $\nu$, with mean $\nu_0=\min\{1, \nu_{UB}/2\}$ and variance $\frac{\nu_0(\nu_{UB}-\nu_0)}{1+\phi_0}$, and
we call $\phi_0$ the precision parameter, whose default value is
$$
\phi_0 = \max\Big\{\frac{\nu_{UB}}{\nu_0}, \frac{\nu_{UB}}{\nu_{UB}-\nu_0}\Big\} + \phi_{inc}.
$$
The parameter $\phi_{inc}$ is an increment to ensure that the prior beta density has boundary 
values equal to zero (where the boundary values are defined either by continuity or by limits). 
The default value of $\phi_{inc}$ is 1. The value of $\phi_{inc}$ can be changed by
changing the argument \code{nu.prec.inc} in the \code{rspde.matern} function. The higher
the value of $\phi_{inc}$ the more informative the prior distribution becomes.

Let us denote a beta distribution with support on $(0,\nu_{UB})$, mean $\mu$ and precision parameter $\phi$ by $\mathcal{B}_{\nu_{UB}}(\mu,\phi)$. If we want $\nu$ to have a prior 
$\nu \sim \mathcal{B}_{\nu_{UB}}($\code{nu_1},\code{prec_1}),
one simply sets \code{prior.nu = list(mean=nu_1, prec=prec_1)}.

Another possibility of prior distribution for $\nu$ is a truncated lognormal distribution. 
Then, we assume that $\log(\nu)$ has prior distributionwith support $(-\infty,\log(\nu_{UB}))$, where $\nu_{UB}$ is the upper bound for $\nu$, with location parameter $\mu_0 =\log(\nu_0)= \log\Big(\min\{1,\nu_{UB}/2\}\Big)$ and scale parameter $\sigma_0 = 1$. More precisely, let $\Phi(\cdot; \mu,\sigma)$ stand for the cumulative distribution function (CDF) of a normal distribution with mean $\mu$ and standard deviation $\sigma$. Then, $\log(\nu)$ has cumulative distribution function given by
$$
F_{\log(\nu)}(x) = \frac{\Phi(x;\mu_0,\sigma_0)}{\Phi(\nu_{UB})},\quad x\leq \nu_{UB},
$$
and $F_{\log(\nu)}(x) = 1$ if $x>\nu_{UB}$. 

To change the prior distribution of $\nu$ to the truncated lognormal distribution, we set the argument \code{prior.nu.dist="lognormal"}. To change these parameters in the prior distribution to, say, \code{m1} and \code{s1}, one sets \code{prior.nu = list(loglocation=m1, logscale=s1)}.

\subsection{Changing the starting values}
The starting values to be used by \pkg{INLA}'s optimization algorithm can be changed by setting one or more of the arguments \code{start.ltau}, \code{start.lkappa} and \code{start.nu} in \code{rspde.matern}.
Here, \code{start.ltau} is the initial value for $\log(\tau)$, \code{start.lkappa} is the inital value for $\log(\kappa)$, and \code{start.nu} is the initial value for $\nu$. 

\subsection{Changing the type of the rational approximation}\label{app:type_rational}         
We have three rational approximations available for obtaining the coefficients in the rational approximation of the function $x^{\alpha}$ on an interval. The BRASIL algorithm \citep{hofreither2021algorithm} and two ``versions''
of the Clenshaw-Lord Chebyshev-Pade algorithm \citep{baker1996pade}, one with lower bound zero and another with the lower bound given in \cite{xiong2022}. The type of rational approximation can, for example, be chosen by setting the \code{type.rational.approx} argument in the \code{rspde.matern} function or by setting \code{type_rational_approximation} in \code{matern.operators}. The BRASIL algorithm corresponds to the choice \code{brasil}, the Clenshaw-Lord Chebyshev pade with zero lower bound and non-zero lower bounds are given, respectively, by the choices \code{chebfun} and \code{chebfunLB}. The type of approximation that is used has an effect on the quality of the approximation, but the choice is seldom of much importance for practical applications. 

\end{appendices}

\bibliography{refs}

\end{document}